Master thesis on Wireless Communications

Universitat Pompeu Fabra

# Application Layer Modeling in Vehicle Networks: Cooperative Maneuver Use Case

Steven Platt

Supervisor: Jesus Alonso-Zarate, PhD

Co-supervisor: Luis Sanabria-Russo, PhD

August 2018

Universitat Pompeu Fabra Barcelona

Page Intentionally Left Blank

# Table of Contents

Dedication

To my partner Gulhan,

who supported me running off to chase a dream.

Page Intentionally Left Blank

## 1 Abstract

In recent years, network function virtualization and a software defined focus has allowed networks to become flexible and extensible in ways not possible previously. Although network modeling tools such as NS-2, NS-3, and OMNet++ have been extended with modules and code to support the absolute latest wireless protocols and medium access standards - there has been a growing gap in simulation of layers above medium access which recent 5G use cases are designed to support. In this thesis, I extend research into the topic of application layer modeling in wireless networks, with focus on upcoming deployments of enhanced vehicle services. In particular, I begin in discussing the evolution of vehicle network standards and use cases, along with the most recent initiative `Project 5GCar`; funded by the European Commission. I go narrower, from the macro trend of vehicular network standards evolution, to compare the state of the art in simulation stacks, designed to support application models in vehicle networks. Within this comparison, simulation environments OVNIS, Veins, iTetris, and VSimRTI are compared for their capabilities. Finally, to measure the qualitative performance of application layer modeling in vehicle networks, I take the cooperative maneuver use case, presented under Project 5GCar; to design an autonomous merge algorithm - completing the steps required to program and model the application in vehicles using the VSimRTI simulation stack.

# 2 Introduction

## 2.1 Project 5GCAR

Funded by the European Commission, the 5GCAR project is a public-private-partnership (PPP) developed to bring together the automotive and communications industries, with research institutions to develop next generation connected vehicle and intelligent transport applications enabled by 5G technologies. In its most recent deliverables, Project 5GCAR outlines five use cases for specific development [16]:

**Cooperative Maneuver:** Sharing of local data for driving intention and trajectory. This data is used to negotiate interaction among groups of vehicles.

**Cooperative Perception:** Sharing of data derived from various sensors. These sensors can be installed to the vehicle, road, or other positions. Development in this area focuses heavily on delivering better than line-of-sight vision to smart vehicles.

**Cooperative Safety:** Sharing of data, optimized for detection of road hazards and safety of other road users, such as pedestrians and cyclists.

**Autonomous Navigation:** Centralized and or distributed processing of maps and routes. These routes are derived from shared vehicle and road network data, and combined with traditional remote sensing and navigation systems.

**Remote Drive:** Controlling a vehicle remotely to enable driving without a physical operator present. This includes actuating all vehicle function remotely, inclusive of braking, steering, and acceleration.

### 2.1.1 Toy Car Lane Merge Model from CTTC

Today, research is being conducted at Centre Tecnològic de Telecomunicacions de Catalunya (CTTC) under the 5GCar project to prove an implementation of the cooperative maneuver use case in connected cars. The hardware implementation under development at CTTC uses a sensing camera to detect vehicle position, while feeding the data to a network controller for processing. Additionally, the network controller, with visibility of the total network state, applies algorithms and issues command and control to vehicles to aid successful navigation and collision avoidance in the cooperative maneuver use case.

Modeling networks on these use cases requires modeling strict vehicle-to-vehicle and vehicle-to-infrastructure communication layers, but also, additional capability to model vehicle mobility, external sensor and camera data, as well as allowing remote interfaces for network controllers and application code to execute real-world command and control during simulations.

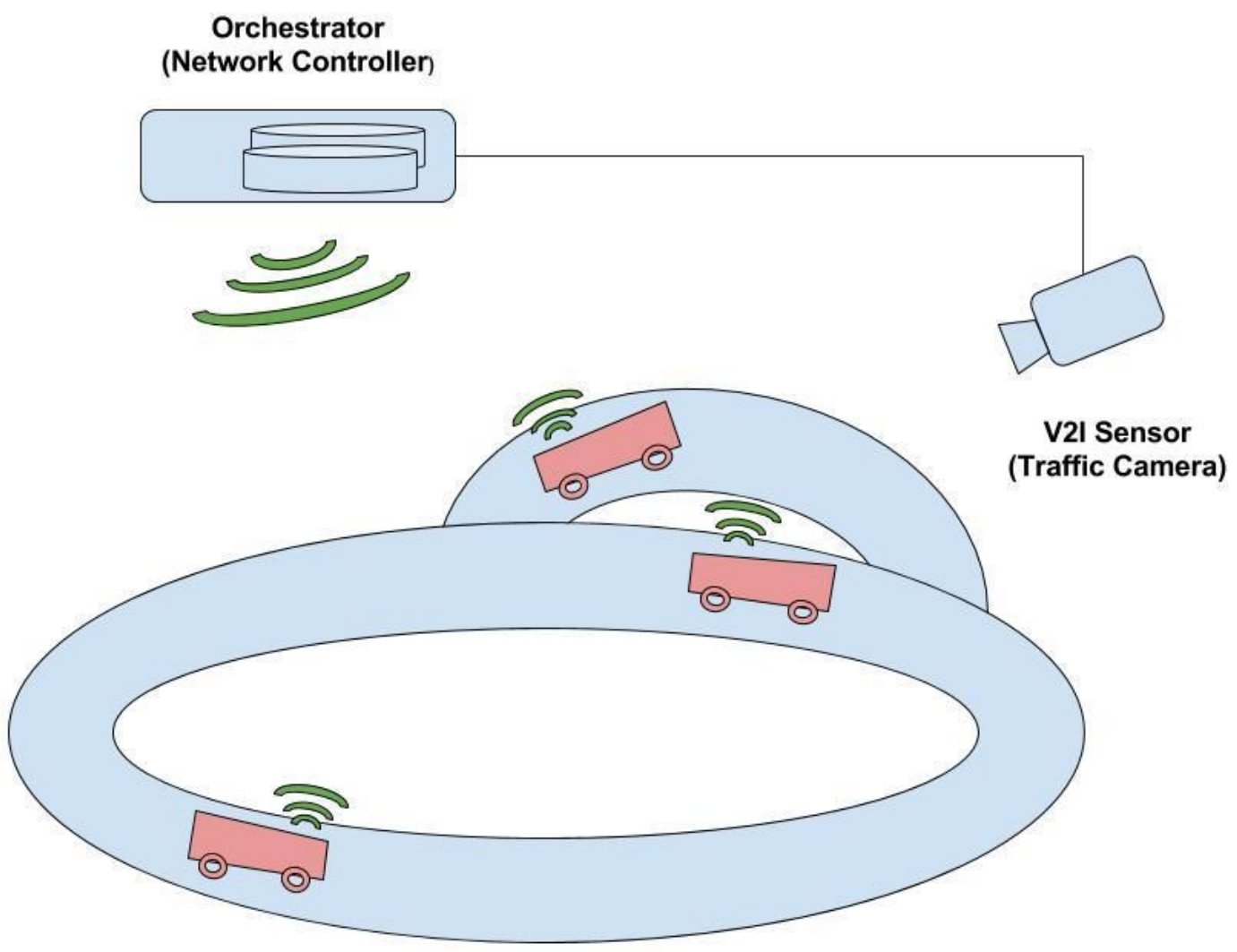

**Figure 1:** Lane merge "Toy Car" model, built by CTTC

### 2.1.2 5GCAR: Cooperative Maneuver Use Case

The toy car model built at CTTC is an example of the Cooperative Maneuver use case, presented in Project 5GCar. Within European Commission requirements outlined for project

5GCAR; vehicles operating in this use case satisfy interaction conditions summarized below [16]:

**Precondition:** Vehicles are in physical proximity to each other, have autonomous driving capabilities, satellite navigation, and are equipped with wireless communications. Participating vehicles are authenticated.

**Triggering Event:** A vehicle wants to join traffic, and merge into a lane.

**Actors:** Traffic vehicle (required), merging vehicle (required), infrastructure sensors (optional), application server (optional). Traffic vehicles must be network enabled, merging vehicles may be network enabled, or not.

Within these conditions, Project 5GCAR intends that wireless communications systems be used to supplement Advanced Driver Assistance Systems (ADAS), which is most commonly implemented using radar, lidar, and vision cameras; to allow better than Line of Sight (LOS) vision and enhanced localization.

## 2.2 Requirements: Application Layer Modeling

Vehicle network standards are still evolving and will receive many edits during the first 5G network deployments. With evolving use cases – there is no single simulation tool today that can model all scenarios and interaction types for vehicle networks. Several existing network simulators have been extended; while new tools have been built to allow connecting data from multiple simulators with custom programming to complete more complex models. This research covers my evaluation of such tools, and progression of choosing to use the VSimRTI for deeper analysis and development of the autonomous merge use case. Although there is a large volume of research and activity surrounding new and existing network standards, there is a much smaller body of research and use case examples for application modeling in vehicle networks. This research looks to focus wholly on the application layer of vehicle network simulation.

# 3 State of the Art

## 3.1 Existing V2X Standards and Use Cases

The development of simulation platforms has followed closely the specifications outlined by standards bodies dating back to 2010. For vehicle network applications, these tentpole standards have been the ETSI Intelligent Transport Systems definition, and later expansions in the form of ITU-R Intelligent Transport Systems, LTE Release 14, and LTE release 15.

### 3.1.1 ETSI Intelligent Transport Systems

In 2010, ETSI defined its first set of Intelligent Transport Systems (ITS) standards [22]. The definition initially outlined six categories and provided examples of intelligent transport application:

**Active Road Safety:** Slow vehicle warning, intersection collision warning, emergency vehicle warning

**Cooperative Traffic Efficiency:** Enhanced route guidance, detour notification

**Cooperative Local Services:** Point of interest notification, media downloading, parking access control

**Global Internet Services:** Fleet management, Insurance services

**Hazardous Location Notification:** Road work warning, emergency brake light notification

**Signage Applications:** In vehicle speed limit display, signal violation (safety)

These categories of Intelligent Transport Systems are one of the earliest uses of the term. Today ETSI is actively updating its standards for ITS, which now include defining applications of vehicle cooperative adaptive cruise control (C-ACC), vehicle platooning, and vulnerable road users (VRU) [23] [24]. These updates are expected to be published as “Release 2”, with the original document being “Release 1”.

### 3.1.2 ITU-R Intelligent Transport Systems

Prior to releases 14 and 15 of the LTE standard by the 3GPP, the United Nations ITU-R published guidance M.1890 (ITUR11-1890) in April 2011 [20]. The document titled "Intelligent Transport Systems - Guidelines and objectives" outlined eight classes of application for network connected transportation systems. These broad categories have since been extended, further defined and implemented in the market in the year after. These eight use classes are:

**Advanced Vehicle Control Systems:** Collision avoidance, enhanced driver vision, pre-crash restraint deployment, and automated road systems.

**Advanced Traffic Management Systems:** Congestion control, traffic control, emissions and parking management.

**Advanced Traveler Information Systems:** Travel guidance, route guidance, and ride matching to facilitate ride sharing services.

**Advanced Public Transportation Systems:** Public transit automation and personalization.

**Advanced Fleet Management:** Vehicle mileage and fuel reporting, international border clearance and safety inspections.

**Emergency Management Systems:** Public safety and emergency notification services.

**Electronic Payment Services:** Electronic payment for toll, parking and other vehicle services.

**Pedestrian Supporting Systems:** Walking directions and vehicle-pedestrian collision avoidance.

### 3.1.3 3GPP LTE Release 14

In 2015, under LTE release 14 technical standards; the 3GPP finalized Technical Specification (TS) 3GPP16-22885, which outlined performance and use requirements for V2X applications in several road conditions. This specification includes speed, distance, latency, and reliability measure focused on driver awareness and warning systems, with reliability requirements as strict as up to 95% [17]. This specification assumed level 1 vehicle communications, using CAM (cooperative awareness message) and DENM (decentralized environmental notification message) messages. The CAM and DENM message have since become a standard format available in a few V2X application simulators, such as VSimRTI. Full detail of TS 22.885 use case specifications are below:

| | Effective Distance | Absolute Speed of UE supporting V2X services | Relative Speed between two UE's supporting V2X services | Maximum Tolerable Latency | Minimum radio layer message reception reliability |
|---|---|---|---|---|---|
| Suburban Major Road | 200 m | 50 km/h | 100 km/h | 100 ms | 90% |
| Freeway/Motorway | 320 m | 160 km/h | 280 km/h | 100 ms | 80% |
| Autobahn | 320 m | 280 km/h | 280 km/h | 100 ms | 80% |
| NLOS/Urban | 150 m | 50 km/h | 100 km/h | 100 ms | 90% |
| Urban Intersection | 50 m | 50 km/h | 100 km/h | 100 ms | 95% |
| Campus / Shopping Area | 50 m | 30 km/h | 30 km/h | 100 ms | 90% |
| Imminent Crash | 20 m | 80 km/h | 160 km/h | 20 ms | 95% |

**Table 1:** 3GPP Release 14 V2X use cases [17]

### 3.1.4 3GPP LTE Release 15

As an enhancement of 3GPP16-22886, the 3GPP later published 3GPP17-22186 as an enhancement for V2X services under LTE release 15. This update allowed more specific requirements among different vehicle network messages, not directly tying them to the environment in which they are deployed. This specification covered five areas [18]:

**Generic Messaging:** General messaging, applicable in all V2X scenarios. Interworking, multi-RAT, and routing messages.

**Vehicle Platooning:** Enabling vehicles to form a platoon and follow each other in a coordinated manner, with a shorter follow gap than usual.

**Advanced Driving:** Exchange of sensing data to increase the perception of connected vehicles.

**Extended Sensors:** Allowing semi-autonomous and fully-autonomous driving, by allowing vehicles to synchronize sensor and trajectory data.

**Remote Driving:** Enables remote driving of vehicles operating in a dangerous environment, or operation of a vehicle for passengers who cannot operate it themselves.

Within each of the above categories, 3GPP17-22186 outlines a similar range of network and use case performance requirements that include: payload size, end-to-end latency, reliability (%), data rate (Mbps), communications range (meters), and transmission rate.

# 3.2 Application Layer Simulators

With development of intelligent transport system applications, a series of simulation platforms began development to allow modeling new software applications in vehicle. There is now a clear split between simulation environments built with focus on communications and mobility modeling (OVNIS and VEINS), and new simulations tacks that have additional programming to allow running vehicle applications on top of existing mobility and communications models (iTetris and VSimRTI).

Depending on the format of simulation, most environment bundle similar components to complete simulations. For platforms focused on communications, this the existing SUMO platform for mobility models, combined with NS-2, NS-3, SWANS or similar environment for communications facilities. Both OVNIS and VEINS use this pairing. Application layer simulators take these components and pair them with additional programming to host external application code, as well as managing data and clock syncing among all connected environments. Comparing performance, based on the above feature requirements, VSimRTI proved most compatible for the intended use. Below is a summary of features and the architecture each framework provides.

**OVNIS**

Developed the University of Luxembourg in 2011, OVNIS is the least feature complete. The project integrate SUMO mobility data and the NS-3 network simulator. OVNIS does support active manipulation of simulation through web sockets, but has not received a code update since 2015, with documentation dating to its original publication in 2011. Mac and Linux platforms are supports for running simulations.

**Veins**

Veins was first released in 2011. It is fully open source under the GPL license and has an active development community. Unlike OVNIS and iTetris; Veins combines SUMO mobility data, with the OmNet++ simulator. which is considered to have a less steep learning curve, compared to NS-3. Veins also has multiple prepackaged extensions to support simulating LTE radio networks as well as shadowing and path loss caused by city buildings – features lacking in both OVNIS and iTetris, without the addition of NS-3 LTE extensions and custom programming. Another important distinction of using Veins, is that it does not provide a fully developed application runtime environment. Instead, users can run any application, as long as the application is written in the C++ language and properly associates to an entity running from within the simulation (vehicle, roadside unit, or similar). Below is a diagram, provided by Veins, to outline the simulators modular functionality.

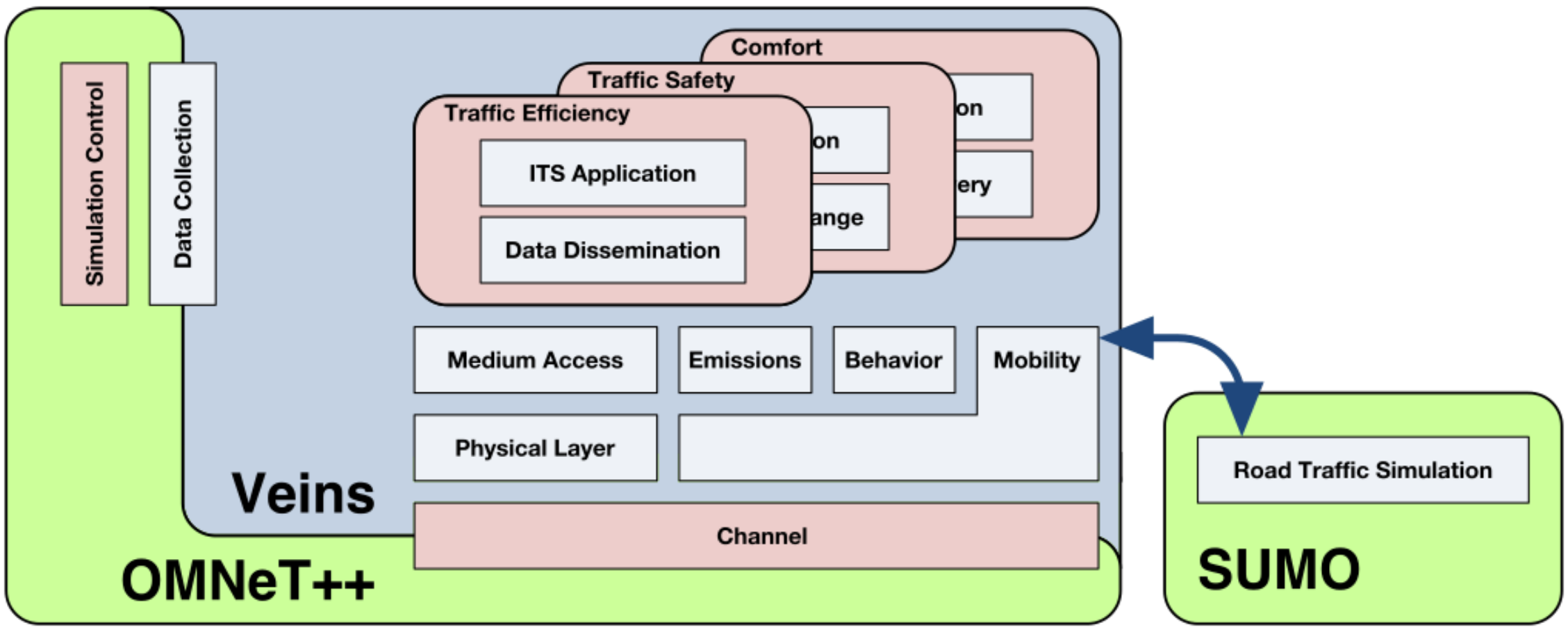

**Figure 2:** Modular Structure of the Veins Simulator [25]

**iTetris**

Released in 2013, iTetris combines, SUMO mobility data with the NS-3 network simulator. iTetris does not include a GUI for visualizing trace and communication data, and support running simulations only on a Linux platform, but the system claims to be language agnostic for its vehicle application framework, and explicitly states support for simulating vehicle applications written in C++, java, and python; a large advantage over VSimRTI, which can only support application written in java. Features and tutorials are tailored for smart transport simulations, such as emissions and congestion models. Another differentiation vs the closed source VSimRTI, all code of the iTetris platform is open source and can be modified and used without special license. Figure 3 shows the structure of the iTetris simulator.

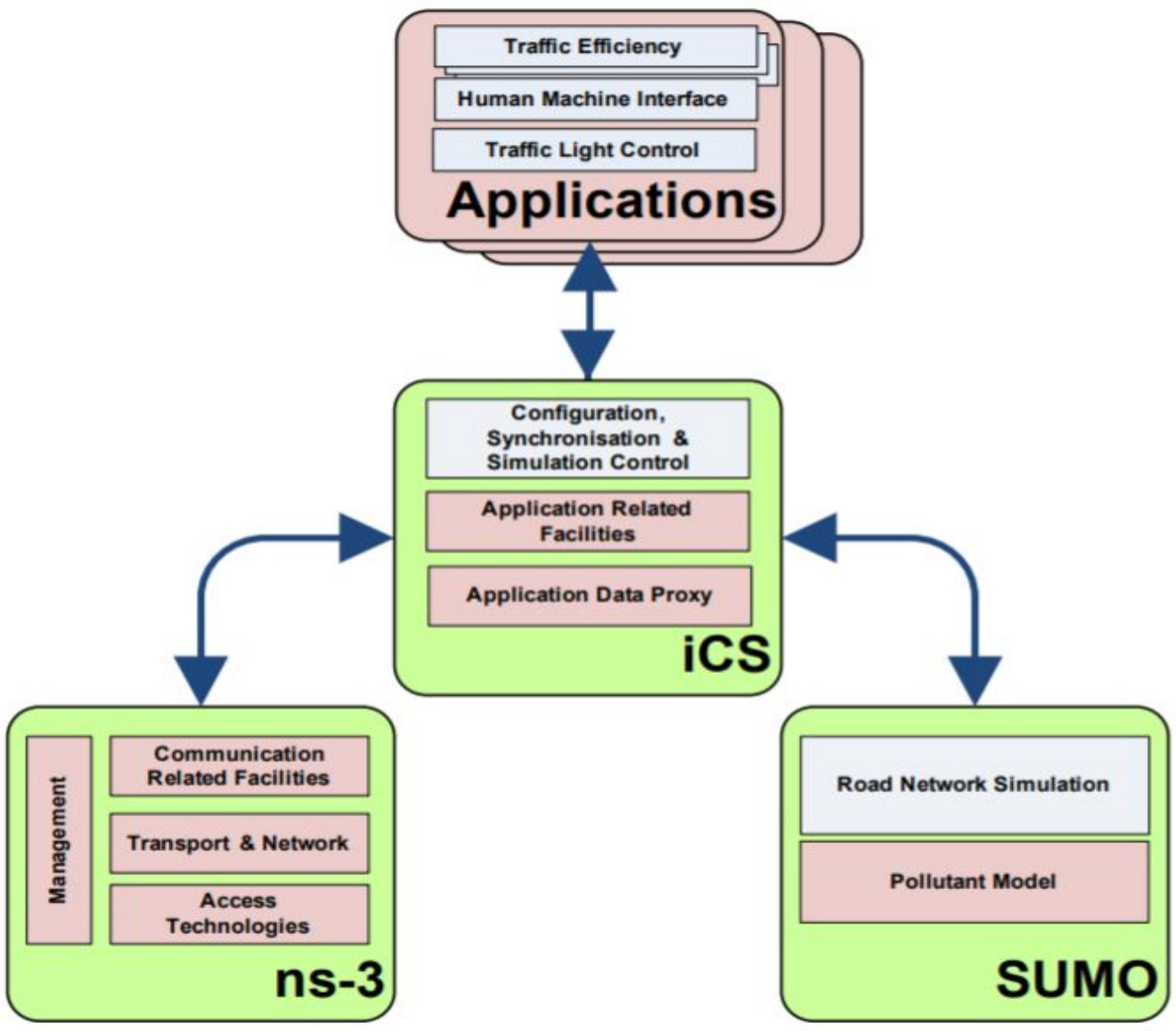

**Figure 3:** Modular Structure of iTetris Simulator [15]

**VSimRTI**

VSimRTI is the newest simulation framework and is also most flexible. It supports SUMO traffic data, combined with NS-3, OmNet++, or SNS network simulators [27]. This is important, since it allows changing the communications layers to suite the experiment or custom code being deployed. The VSimRTI framework is also best at visualizing data with support for Open Street Maps traffic overlays. VSimRTI supports Windows, Mac, and Linux. Unlike the other simulators listed, VSimRTI is not open source, and running the simulator requires a physical license by granted by the Fraunhofer institute, who supports the simulator. VSimRTI relies on a few open source projects for the core of it simulation environment, but a large portion of the simulators code is proprietary and encompasses application code written to simulate vehicle and roadside unit operating systems. This includes items such as minimum following distance, how quickly vehicle speed is adjusted, and providing GPS, messaging, and other application functionality to entities in simulation. Figure 4 shows the VSimRTI runtime infrastructure and the open source components on which it is based.

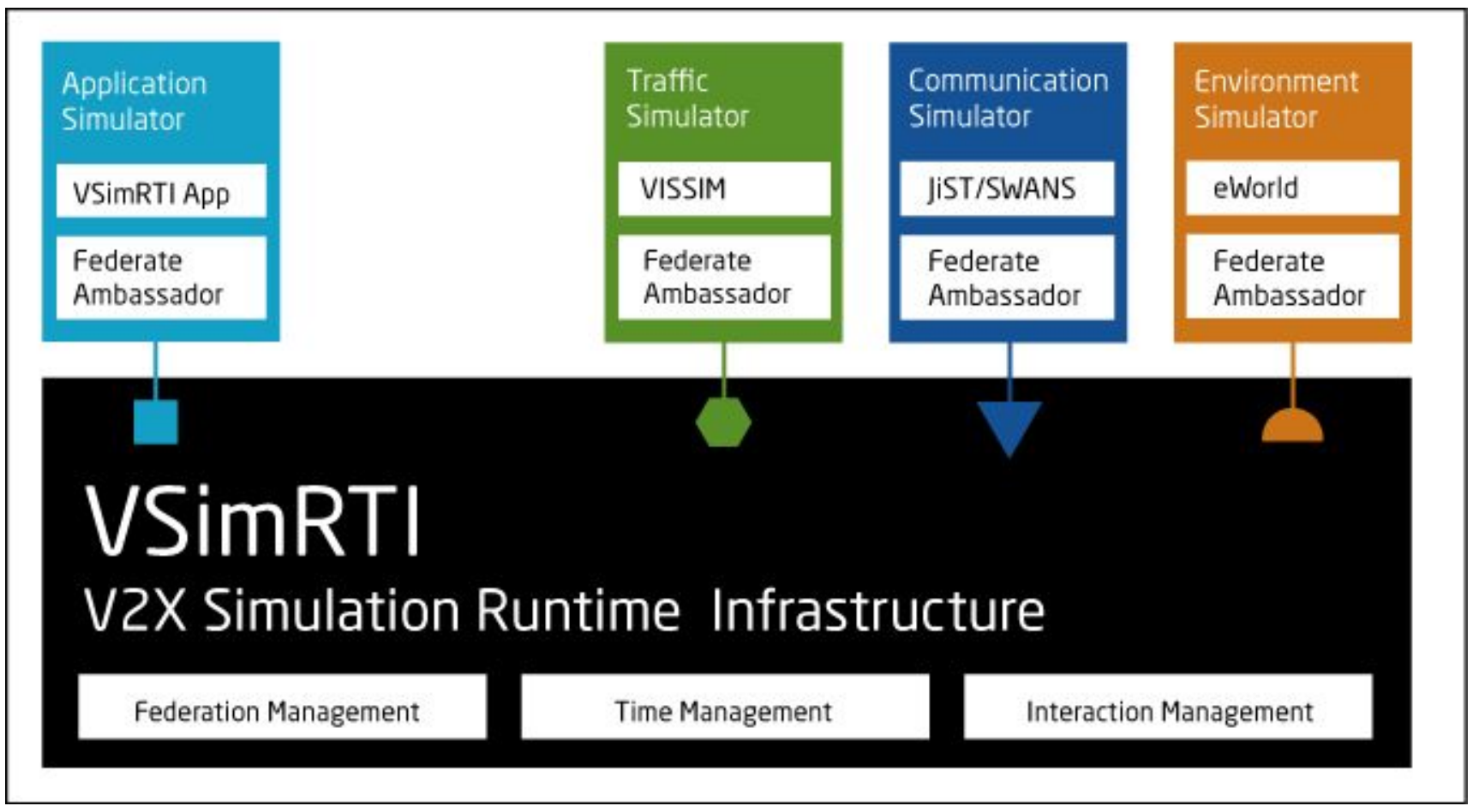

**Figure 4:** Modular Structure of VSimRTI Simulator [26]

# 4 Proposed Solution

## 4.1 The VSimRTI Simulator

Based on planned research, VSimRTI is selected as the simulation framework. It has the best native support for visualizing data while being competitive to deliver all other layers required in the simulation stack. For maximum compatibility Ubuntu Linux 16.04 LTS is chosen as the installation base for all components.

### 4.1.1 Obtaining a License

VSimRTI requires a license to operate. When downloading the VSimRTI application, running the "firstStart.sh" from its root folder generates a text file with system info, including CPU core, operating system, and RAM installed. After emailing this file to the VSimRTI email distribution (vsimrti@fokus.fraunhofer.de), a username is issued with an assigned expiry date – the expiry can be extended with an additional email request.

**Note:** Running the firstStart.sh file requires root access (sudo) the java runtime. This is not installed by default on Ubuntu 16.04 LTS, I installed it using commands:

```
#add the openjdk repository and install open jdk
sudo add-apt-repository ppa:openjdk-r/ppa
sudo apt-get update
sudo apt-get install openjdk-8-jdk

#add linux environment variable for java installation
export JAVA_HOME=/usr/lib/jvm/java-8-openjdk
```

Beyond validating the required license, no further installation is required of VSimRTI at this time.

### 4.1.2 Installing SUMO

Installing SUMO is the simplest portion of the build, as it is included in the stock Ubuntu 16.04 repository. It is installed with:

```
sudo add-apt-repository ppa:sumo/stable
sudo apt-get update
sudo apt-get install sumo sumo-tools sumo-doc
which SUMO #confirms install and shows installation directory
```

## 4.2 Autonomous Merge Algorithm

With the simulation environment chosen, this section covers the details of the algorithm planned for managing lane merges of approaching vehicles in a custom-built simulation scenario, named “Castelldefels”.

The Castelldefels simulation assumes 802.11p communications, rather than a cellular network. This allows the same simulation to be built with physical test hardware or expanded by the CTTC if required.

### 4.2.1 Algorithm Message Flow

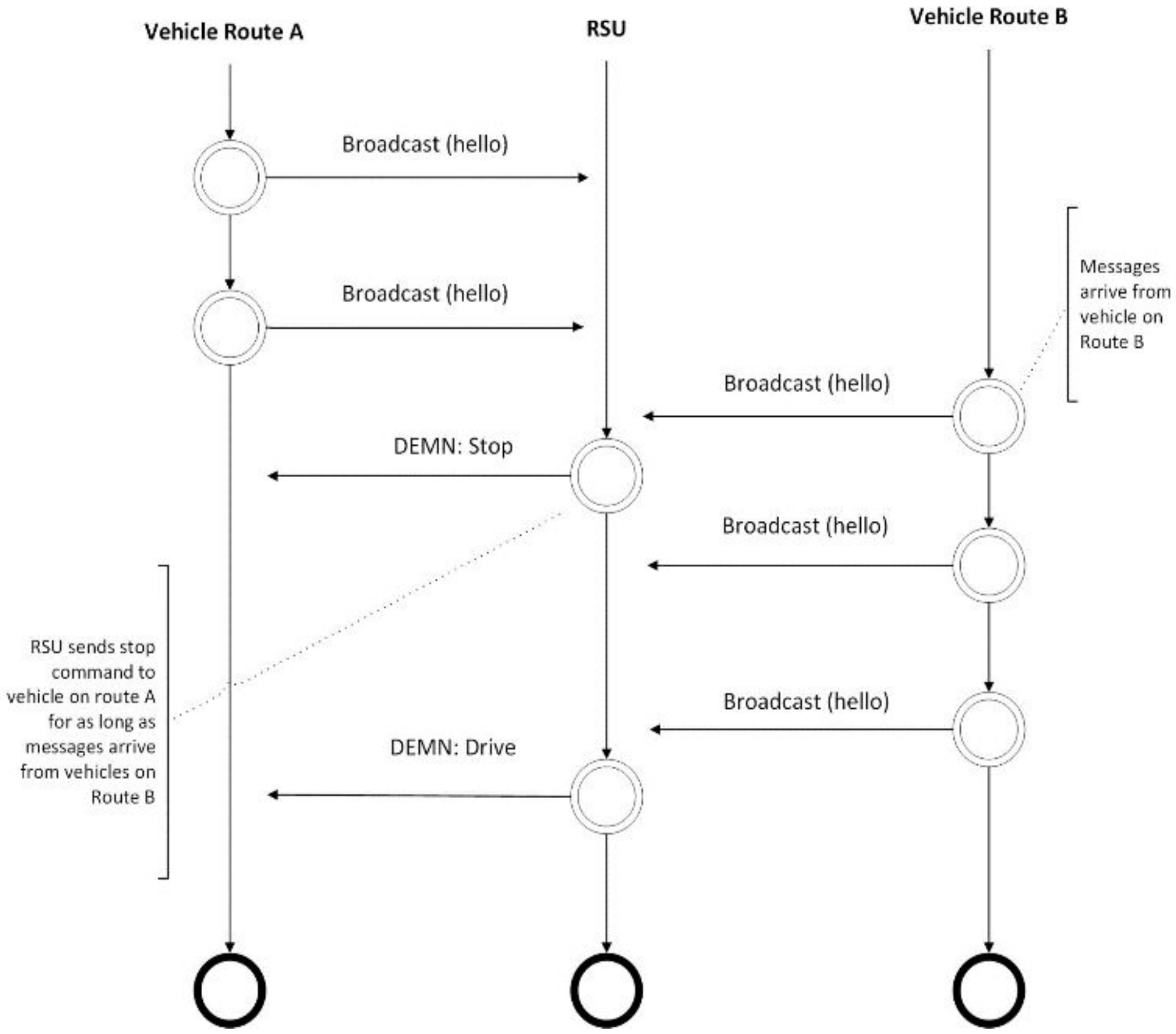

**Figure 5:** Message flow between the roadside unit and merging vehicles

Based on vehicle routes calculated through SUMO, the expectation is that a roadside unit at the junction point will sit listening always for oncoming vehicles. Vehicles in the simulation are limited to two predefined routes – I refer to them as routes "A", and "B". All vehicles in the simulation are broadcasting beacons using 802.11p as they drive the route. Once in close range, the roadside unit will begin receiving these broadcasts.

The algorithm gives priority to vehicles entering on route B. Vehicles on route A pass and enter the roundabout unrestricted, until messages arrive to the RSU from vehicles in range on route B. After the first message arrives – the RSU sends a control command to the vehicle on route A to 'stop', until it no longer receives the broadcast beacons from the vehicle on route B. At that time, the RSU send a second control command to the route A vehicle to 'drive'.

### 4.2.2 Algorithm Logic

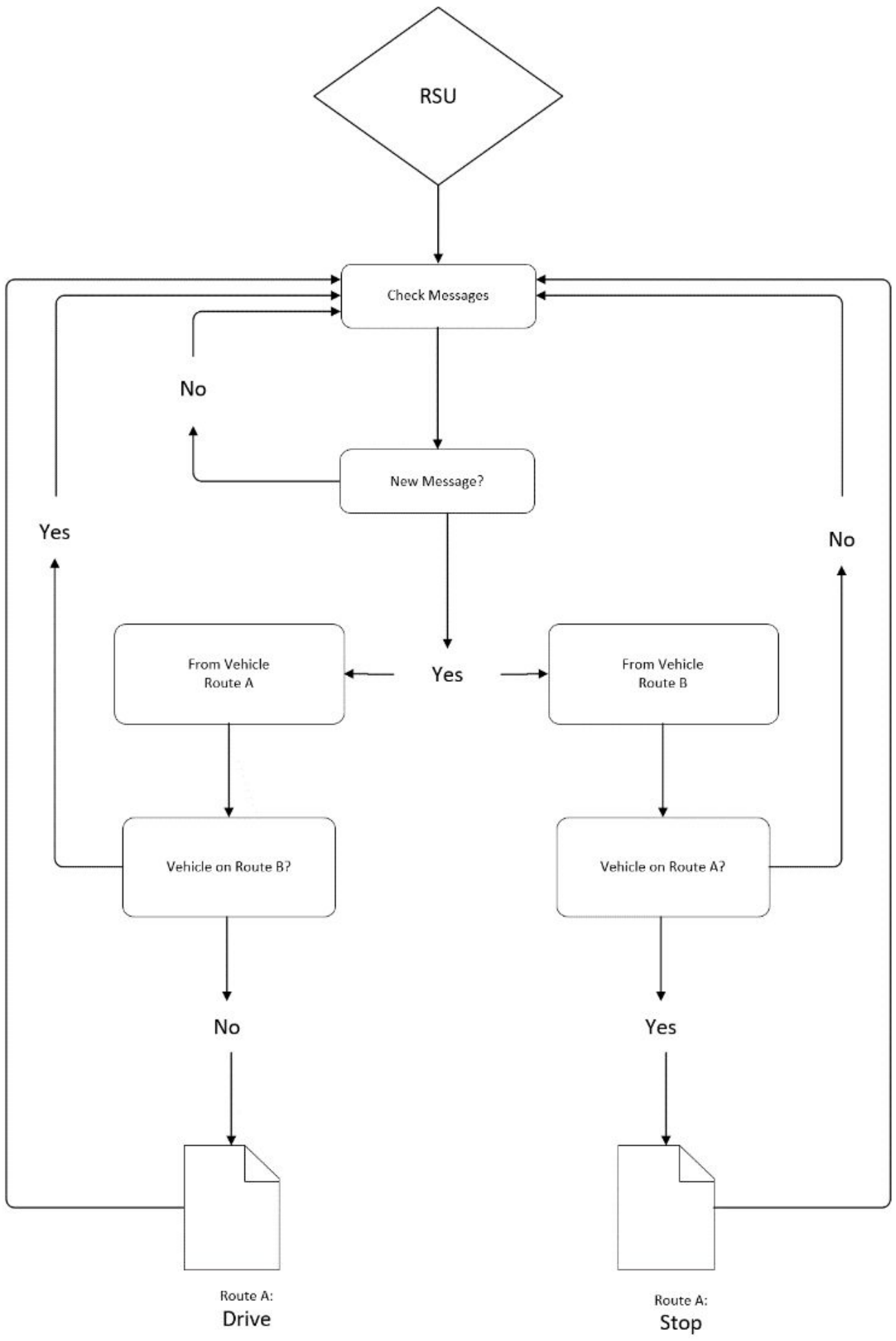

The above image shows the full detail of the algorithm. As part of the vehicle operating system implemented by VSimRTI, the vehicles are deployed with controls that manage the acceleration, top speed, and minimum follow distance of vehicles. They are essentially driving autonomously in a platoon, along the simulation routes, until the algorithm controls are applied. For this reason, no controls or settings relating to these items are included in the algorithm. I can limit it only to the management of the merge action.

Note that the algorithm code runs within the 'Castelldefels RSU' java class on the roadside unit. For additional context I represent again the logic in written form below:

**Autonomous Merge Algorithm: Written Form**

1. Function: Castelldefels RSU ( $M_B$ , $M_Q$ , $T$ , $M_R$ , $V^N$ , $R$ , $P$ )

**Input:**

$M_B$ is the beacon message being sent from all vehicles instantiated for the simulation. If it is null, there are no vehicles within 802.11p broadcast range of the roadside unit.

$M_Q$ is a queued message. Messages are held for 5 seconds before they are discarded.

$T$ represents the time delay before a beacon that was queued is emptied.

$M_R$ is the message response sent by the roadside unit, as reaction to broadcasting vehicles being within range. $M_R = 1$ if drive is desired, and 0 to stop a vehicle.

$V^N$ represents the unique vehicle negotiating a merge. There are $N^{th}$ vehicles participating in a given simulation run.

$R$ is the roadside unit coordinating the network assisted vehicle merge.

$P$ is the path on which the vehicle is traveling, this path is the SUMO generated road ID, and is variable mapped as either route A or B in the simulation scenario.

**Output:** *Drive* or *Stop*

2. **if** ( $M_B \neq null$ & $P$ = A & $M_Q = null \vee$ A) **then**

3. $M_R = Drive$;

4. **else**

5. $M_R = Stop$;

6. return $M_R$ ;

### 4.2.3 Environment Map

Applying the autonomous merge algorithm in virtual space, figure 7 shows the layout of the Castelldefels simulation. Note that vehicles in the diagram are not to scale. The diagram shows physical representation of vehicles concurrently arriving on from opposing routes A and B. The controlling roadside unit is also displayed. The blue circle surrounding the center of the diagram represents the communications range in which the roadside unit receive ad hoc beacons from arriving vehicles. This assumes a standard transmit and received powers along with uniform path loss.

Beyond these three actors in the environment, I show representation of the beacon data flow, from the vehicles to the tower, and corresponding stop and drive response sent in return. Last, there is the target merge zone, the roundabout, marked with a blue triangle.

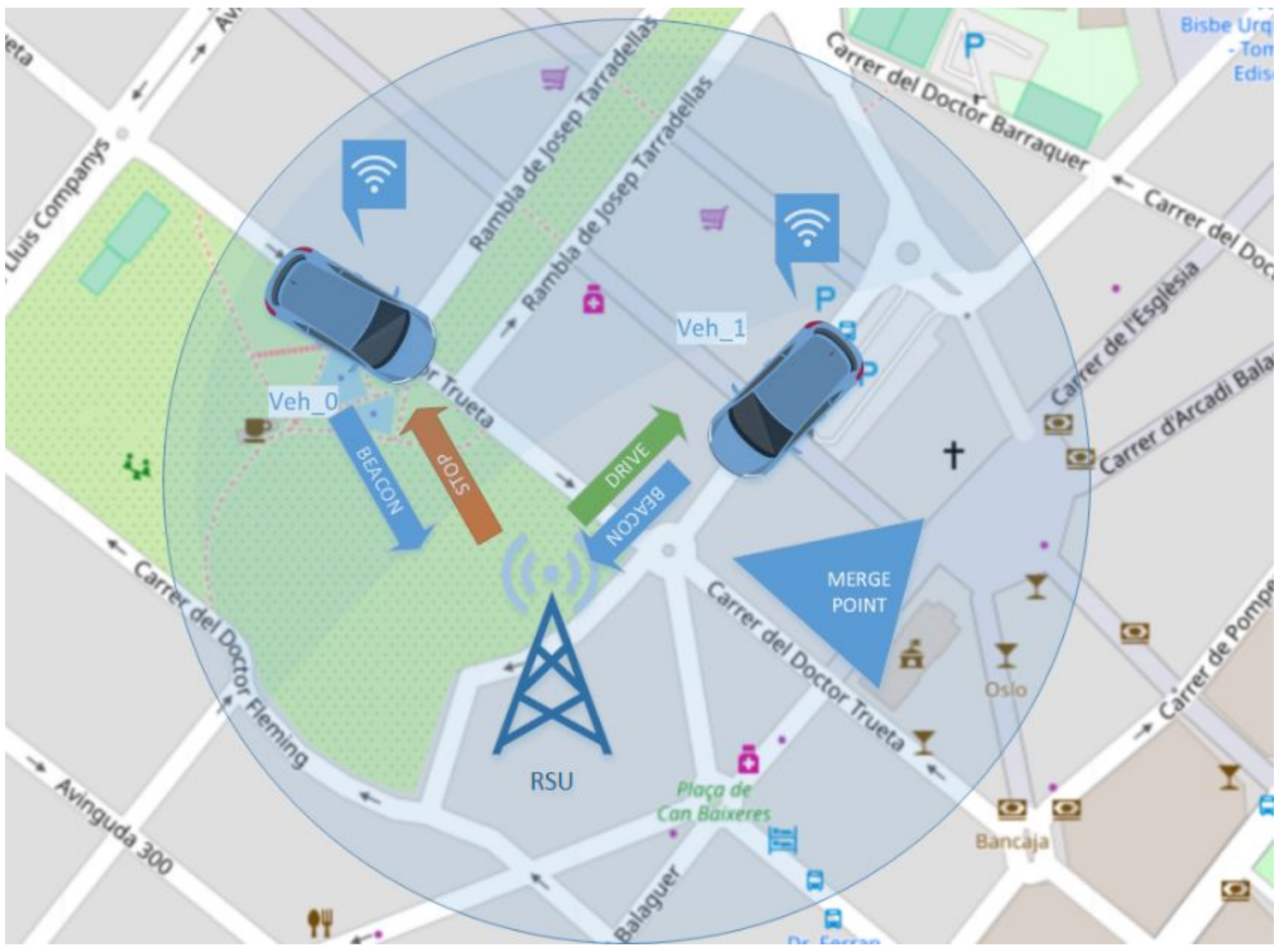

**Figure 7:** Environment map of participants in the Castelldefels simulations

### 4.2.4 Application Java Class Map

Table 2 shows which java application classes are running on each actor in the simulation. All vehicle and all roadside units are assumed identical in the simulation.

The roadside unit has the simpler structure of the entities in the environment. Its programming is being housed in a single Java class, named “Castelldefels RSU”. This coding equips the roadside unit with 802.11p ad-hoc network abilities, along with the algorithm logic for receiving vehicle beacon data, parsing its contents, and responding with the “Stop” or “Drive” commands shown in the previous figure 7.

Programming on the vehicle includes two java classes, these are the ‘Slow Down App’, and ‘Network Merge Assist’ classes. The Network Merge Assist class works similar to the Castelldefels RSU application class, it equips the vehicle with 802.11p networking, and manages the queuing and broadcast of vehicle data, such as location and route, while listening for responses from a roadside unit, and when the message is received, it parses the Stop or Drive message. If a Stop message is received, this class calls the Slow Down App, which brings the vehicle to a stop using internal vehicle control messages.

| **Java Application Classes** | |
|---|---|
| **Roadside Unit** | **Vehicle** |
| Castelldefels RSU Class | Network Merge Assist Class<br>Slow Down App Class |

**Table 2:** Mapping of Java application classes to simulation entities

### 4.2.5 Autonomous Merge Fallback

VSimRTI has thorough documentation of its code capabilities, Java API, and provides explicit code samples when downloading the simulator. In reading the provided VSimRTI documentation, the Java API follows very closely the DENM and CAM messaging standards outlined in LTE release 14 and may have limits to the amount and types of data these messages can carry between vehicles in simulation. Because source code for VSimRTI message handling is not open source - I have added additional fallback programming, based on known functions demonstrated in the Barnim simulation bundled with VSimRTI. In this scenario, road sensors are used in place of direct message exchange between vehicles and RSU. For the Castelldefels simulation implementation, this means a sensor area covering the area of the roundabout signals a speed event to all oncoming vehicles, to force a speed reduction as they navigate the known hazardous merge. Coverage for this sensor notification area is set equal to that shown in figure 7. Code for the fallback sensor event is covered in later section “Signaling a Sensor Event”, with additional coding contained in the java class “Network Merge Assist” running on simulated vehicles.

With the algorithm logic, environment, application classes and fallback functions mapped; I move onto building the simulation within VSimRTI.

# 5 Development

## 5.1 Castelldefels Simulation: Configuring VSimRTI

To unpack the capabilities and shortcomings of the VSimRTI simulator, I created a simulation scenario named "Castelldefels", as it takes place in the town of Castelldefels, Spain.

VSimRTI refers to simulations as 'scenarios', and require each scenario have a unique folder containing a number of configuration files that are specific to VSimRTI. Figure 8 shows an outline of the ten configuration folders used by a VSimRTI simulation. Of the ten configuration folders, SNS, Cell2, Battery, and Visualizer folders require no modification for the Castelldefels simulation and are left as defaults, copied from the 'Barnim' simulation example provided with the VSimRTI simulator.

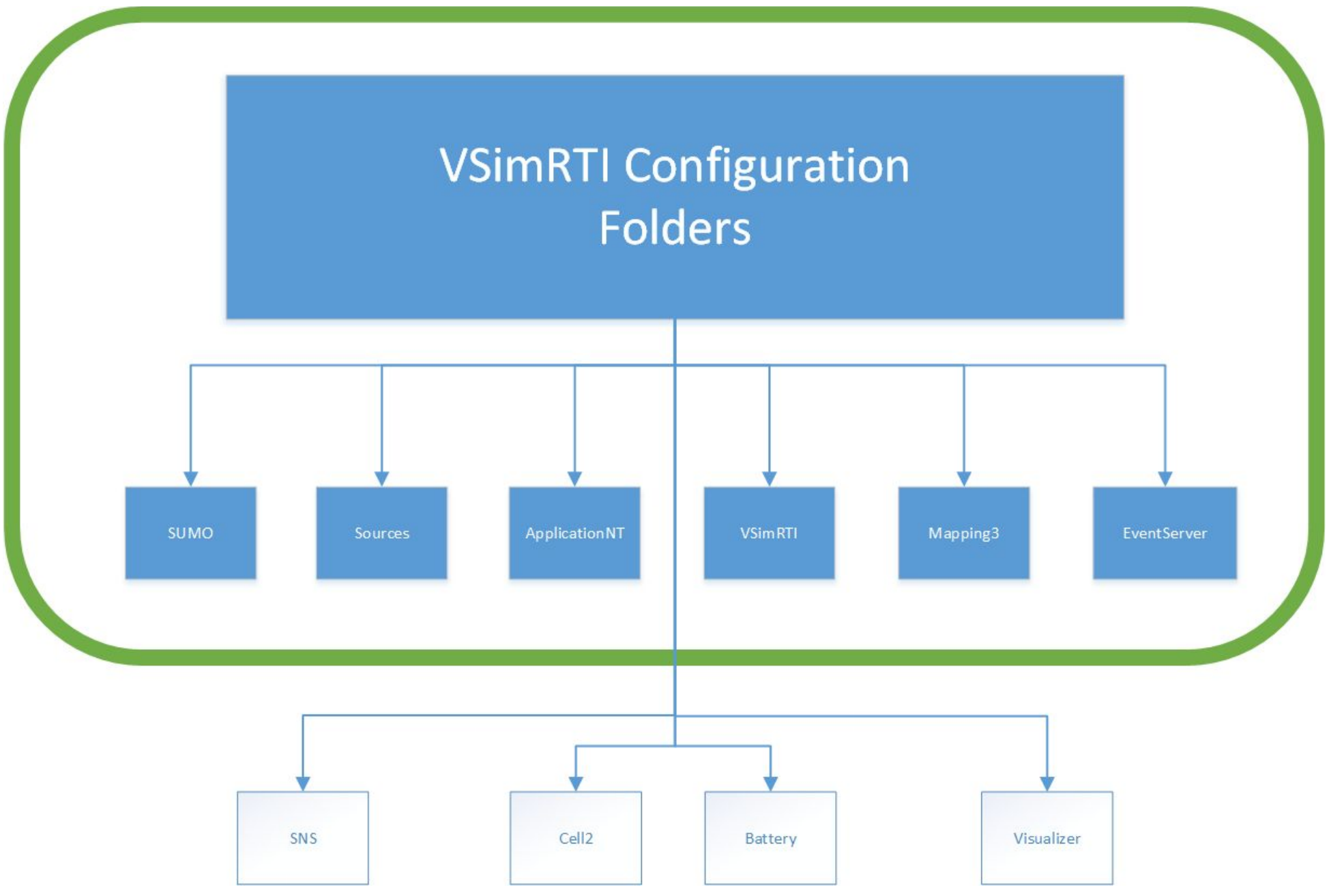

**Figure 8:** VSimRTI Configuration Folders

### 5.1.1 Importing OpenStreetMap Data

For the simulation, I download the map region for Castelldefels, Spain using the "export" option at the openstreetmaps.org website. The resulting file has a ".osm" file extension.

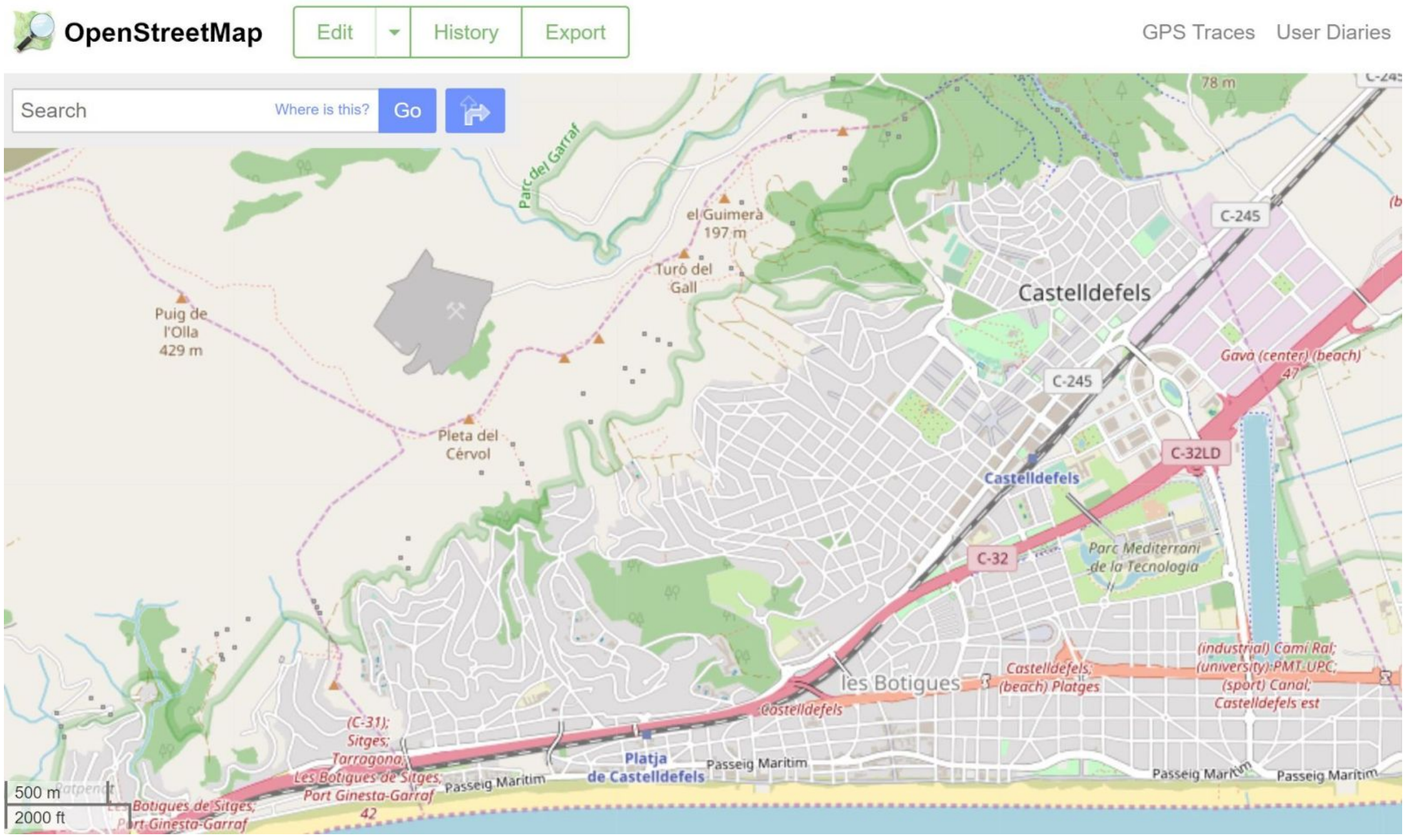

**Figure 9:** OpenStreetMap view of Castelldefels, Spain

#### Convert the OpenStreetMap file to a VSimRTI DB file

Data embedded into the .osm file is not naively readable by either SUMO or VSimRTI. The provided java application "scenario-convert" is used to extract the .osm road network data and place it into a VSimRTI database file:

```
java - jar scenario - convert -17.1. jar -- osm2db -i path / to / source . osm -o
```

#### Export DB Map Data to SUMO

Using the newly generated .db file from the previous step – I export the data to files understandable by the SUMO mobility simulator:

```
java - jar scenario - convert -17.1. jar -- db2sumo -d path / to / database . db
-n
```

When completed, this creates a series of new .xml files for use with SUMO. These files are placed into a new "sumo" folder, and a new '.sumo_config.json' is created, to tell SUMO which .sumo.cfg file it should use for configuration:

```
{
    "sumoConfigurationFile" : "castelldefels.sumo.cfg"
}
```

**Generate SUMO Routes**

After exporting the DB map data to SUMO, I generate traffic routes. This requires the latitude/longitude of the start and end points. When complete, a new route file in .xml format is generated. Because the Castelldefels simulation is not intended to make use of traffic light controls, the additional switch statement required to export traffic light data to SUMO are not included. The final files are then placed in the 'sumo' folder with the .xml files previously generated.

```
java -jar scenario-convert-17.1.jar -d path/to/database.db -g
--route-begin-latlon LAT,LON --route-end-latlon LAT,LON
```

### 5.1.2 Storing Code and SUMO Data with ApplicationNT

For the simulation to work, the created .db files should be placed in a newly created ApplicationNT folder. More importantly, this folder holds the final exported .jar files containing all java classes and programs that will run on entities existing in the simulation. No configuration is required in the ApplicationNT folder, rather it serves only to store these relevant files used in simulation.

### 5.1.3 VSimRTI Configuration File

The final update is to the VSimRTI config file in the scenario folder. This file issues the center latitude/longitude position of the visualization, the Cartesian offset (explained in the VSimRTI manual), the IP address pools, and settings for which simulator components are active during the scenario.

```
<?xml version="1.0" encoding="UTF-8"?>
<configuration xmlns:xsi="http://www.w3.org/2001/XMLSchema-instance"

xsi:noNamespaceSchemaLocation="https://www.dcaiti.tu-berlin.de/research/simulatio
n/download/get/scenarios/scenarioname/vsimrti/vsimrti_config.xsd">
    <simulation>
        <!-- Scenario name -->
        <id>Castelldefels</id>
        <!-- Simulation time frame -->
        <starttime>0</starttime>
        <endtime>1200</endtime>
        <!-- Seed for pseudo random number generator used by most of the
federates -->
        <randomSeed>268965854</randomSeed>
        <!-- Projection between global and cartesian coordinates -->
        <!-- centerCoordinates: roughly the center of simulation area -->
        <!-- cartesianOffset: can be found in the generated network file for the
traffic simulator, e.g. the .net.xml file for sumo -->
        <WGS84UTMTransformConfig>
        {
            "centerCoordinates": {
                "longitude": 1.58,
                "latitude": 41.16
            },
            "cartesianOffset": {
                "x": -411320.95,
                "y": -4568671.76
            }
        }
        </WGS84UTMTransformConfig>
        <!-- Global IP management -->
        <IPResolverConfig>
        {
            netMask: "255.255.0.0",
            vehicleNet: "10.1.0.0",
            rsuNet: "10.2.0.0",
```

```
            tlNet: "10.3.0.0",
            csNet: "10.4.0.0"
        }
        </IPResolverConfig>
        <threads>1</threads>
    </simulation>
    <federates>
        <!-- Cellular network simulator -->
        <federate id="cell2" active="false"/>
        <!-- V2X (ad hoc) network simulators -->
        <federate id="omnetpp" active="false"/>
        <federate id="ns3" active="false"/>
        <federate id="sns" active="true"/>
        <!-- Traffic simulators -->
        <federate id="sumo" active="true"/>
        <!-- Application simulator -->
        <federate id="applicationNT" active="true"/>
        <!-- Environment simulator -->
        <federate id="eventserver" active="true"/>
        <!-- Electric mobility simulators -->
        <federate id="battery" active="false"/>
        <federate id="chargingstation" active="false"/>
        <!-- Mapping -->
        <federate id="mapping3" active="true"/>
        <!-- Visualization -->
        <federate id="visualizer" active="true"/>
    </federates>
</configuration>
```

### 5.1.4 Finalizing Vehicle Routes

Earlier, I outline the process of creating the base Castelldefels scenario. The bulk of the work at that time was to import the OpenStreetMap map data, and build the SUMO routes.

```
java -jar scenario-convert-17.1.jar -d path/to/database.db -g
--route-begin-latlon LAT,LON --route-end-latlon LAT,LON
```

The stock command above is from the VSimRTI manual. When using this command, an additional command switch can also be used to specify the number of routes generate. There is often more than one way to get between two points on a map. In my case, I chose to generate two routes.

Generating two routes creates the 'T-junction' needed to simulate the network control of cross traffic. When using the VSimRTI provided 'scenario-convert' program – SUMO does not allow fine control of which route it maps, and the output of the SUMO route is a string of coordinates – so this can't be easily manipulated after. In other uses, the most reliable way to force a route, would be to generate a dozen of more routes with SUMO, then choose a route after that matches the desired flow. This applies specifically for using SUMO with VSimRTI.

### 5.1.5 Mapping Routes to Vehicles

The mapping 3 folder contains a .json file that maps the java classes stored in ApplicationNT, to entities within the simulation. This also includes configuration of the number of vehicles, vehicle speed, follow distance, acceleration speed, as well as the latitude/longitude location of the transmitting cell tower. If an entity does anything during simulation – it must be mapped in the mapping 3 configuration. After routes were generated with 'scenario-convert', I map them to vehicles, to see cars drive along the route in the running simulation. My updated mapping_config.json configuration is below:

```
{
    "prototypes":[
        {
            "applications": [
"com.telecomsteve.vsimrti.applications.VehicleBeacon",
"com.telecomsteve.vsimrti.applications.SlowDownApp",
"com.telecomsteve.vsimrti.applications.NetworkMergeAssist"
            ],
            "name":"PKW",
            "accel":2.6,
            "decel":4.5,
            "length":5.00,
            "maxSpeed":20.0,
            "minGap":2.5,
```

```
            "sigma":0.5,
            "tau":1
        },
        {
"applications":["com.telecomsteve.vsimrti.applications.CastelldefelsRSU"],
           "name":"RSU"
        }
     ],
    "rsus":[
        {
           "lat":41.28039945484303,
           "lon":1.975863217521691,
           "name":"RSU"
        }
    ],
     "vehicles":[
         {
            "startingTime": 5.0,
            "targetDensity":500,
            "maxNumberVehicles": 100,
            "route":"1",
            "types":[
                {
                    "name":"PKW"
                }
            ]
        },
        {
            "startingTime": 5.0,
            "targetDensity":120,
            "maxNumberVehicles": 20,
            "route":"2",
```

```
            "types":[
                {
                    "name":"PKW"
                }
]}]}
```

Note that vehicles are split as json objects, with some mapped with "route 1", and a second mapped with "route 2". I also map the empty java classes to vehicles and RSU in the simulation.

When running the simulation, I get two paths of cars, routing from source to destination. Communications and network events are not occurring yet.

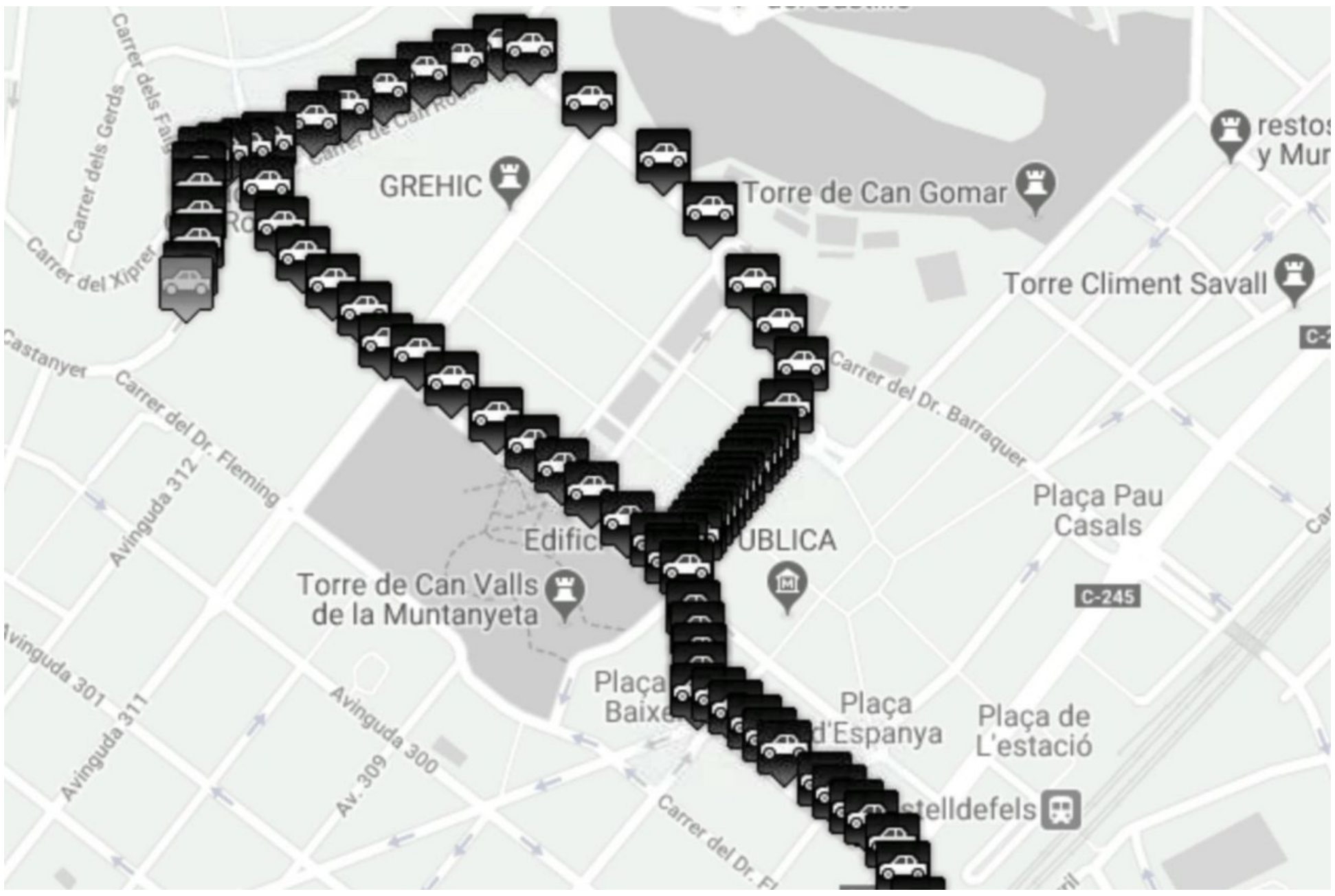

**Figure 10:** Castelldefels vehicle routes, without communications

### 5.1.6 Signaling a Sensor Event

If messaging for the autonomous merge algorithm is not working properly, a target zone that covers the roundabout is set to trigger a sensor event for oncoming vehicles. Reading and acting upon sensor data is handled by the empty Network Merge Assist java class running on the vehicle. The details of the event that is triggered however is managed by the Event Server config in VSimRTI. To trigger a sensored speed event, the 'eventserver_config.json' file is updated within the 'eventserver' folder of the scenario to reflect the Castelldefels target coordinates below.

```
{
    "events" : [
        {
            "type" : "Speed",
            "rectangle" : {
                "a" : {
                    "latitude" : 41.28105,
                    "longitude" : 1.97384
                },
                "b" : {
                    "latitude" : 41.27961,
                    "longitude" : 1.97701
                }
            },
            "strength" : 1,
            "time" : {
                "start" : 0,
                "end" : 1220
            }
        }
    ]
}
```

# 5.2 Autonomous Merge Application Programming

## 5.2.1 Castelldefels Roadside Unit Programming

The Castelldefels Roadside Unit contains a single java class name CastelldefelsRSU. This class extends the RoadsideUnitOperatingSystem class, and implements the VehicleApplication, CommunicationsApplications, and RoadsideUnitApplication classes. which causes it to inherit several features and functions stored in the proprietary code of VSimRTI.

Outside of declaring the above classes, the CastelldefellsRSU class also imports twenty-six additional classes of proprietary VSimRTI code to handle functions of instantiating 802.11p wireless, GPS, as well as handling V2X and DENM message formats. The total list of dependencies is omitted here for brevity but are included for all java classes in aggregate in the results section.

At this stage of development, functions for calculating priority vehicle routes and returning drive / stop messaging are omitted. These code blocks trigger several dependency errors relating to the proprietary code of VSimRTI and are detailed in the results section. For now, the RSU is equipment with 802.11p connectivity and the ability to receive and send V2X messaging, as well as signal hazard events over the 802.11p medium.

**CastelldefelsRSU Class**

```
package com.telecomsteve.vsimrti.applications;

import ; /*import statements removed for brevity*/


public class CastelldefelsRSU

 extends AbstractApplication<RoadSideUnitOperatingSystem>

 implements VehicleApplication, CommunicationApplication, RoadSideUnitApplication

{

 private static final int MAX_ID = 5000;

 private static final long INTERVAL = 2000000000L;

 private static final GeoPoint HAZARD_LOCATION = GeoPoint.latlon(52.633047D,
13.565314D);

 private static final String HAZARD_ROAD = "-3366_2026362940_1313885502";


 public CastelldefelsRSU() {}

 private DENM constructDENM()

 {
```

```
   TopocastDestinationAddress tda =
TopocastDestinationAddress.getBroadcastSingleHop();

   DestinationAddressContainer dac =
DestinationAddressContainer.createTopocastDestinationAddressAdHoc(tda,
AdHocChannel.CCH);

   MessageRouting routing = new MessageRouting(dac,
((RoadSideUnitOperatingSystem)getOperatingSystem()).generateSourceAddressContaine
r());

   int strength =
((RoadSideUnitOperatingSystem)getOperatingSystem()).getStateOfEnvironmentSensor(S
ensorType.Speed);

   return new DENM(routing, new
DENMContent(((RoadSideUnitOperatingSystem)getOperatingSystem()).getSimulationTime
(), HAZARD_LOCATION, "-3366_2026362940_1313885502",

     SensorType.Speed, strength, WeatherWarningApp.getNewSpeed(SensorType.Speed),
0.0F));

 }

  private void sample() {

   DENM denm = constructDENM();


((RoadSideUnitOperatingSystem)getOperatingSystem()).getAdHocModule().sendV2XMessa
ge(denm);

   getLog().infoSimTime(this, "Sent DENM vehicle message", new Object[0]);


((RoadSideUnitOperatingSystem)getOperatingSystem()).getEventManager().addEvent(ne
w Event(((RoadSideUnitOperatingSystem)getOperatingSystem()).getSimulationTime() +
2000000000L, this)); }


 public void setUp() {

   getLog().infoSimTime(this, "Starting road side unit application", new
Object[0]);
```

```
((RoadSideUnitOperatingSystem)getOperatingSystem()).getAdHocModule().enable(new
AdHocModuleConfiguration()
     .addRadio()
     .channel(AdHocChannel.CCH)
     .power(50)
     .create());
   getLog().infoSimTime(this, "Activating 802.11p AdHoc WiFi Module", new
Object[0]);
 }
  public void afterUpdateVehicleInfo()
 {
   List<? extends Application> applications =
((RoadSideUnitOperatingSystem)getOperatingSystem()).getApplications();
   IntraVehicleMsg message = new
IntraVehicleMsg(((RoadSideUnitOperatingSystem)getOperatingSystem()).getId(),
getRandom().nextInt(0, 5000));

   for (Application application : applications) {
     Event event = new
Event(((RoadSideUnitOperatingSystem)getOperatingSystem()).getSimulationTime() +
10L, application, message);

((RoadSideUnitOperatingSystem)getOperatingSystem()).getEventManager().addEvent(ev
ent);
   }}
  public void receiveV2XMessage(ReceivedV2XMessage receivedV2XMessage)
 {
   getLog().infoSimTime(this, "Received message from {}", new Object[] {
receivedV2XMessage.getMessage().getRouting().getSourceAddressContainer().getSourc
eName() });
```

```
   getLog().infoSimTime(this, "Vehicle position confirmed as: {}", new Object[] {
receivedV2XMessage.getMessage() });

   sample(); }

   public void processEvent(Event event)

   throws Exception

 {}

 public void beforeUpdateConnection() {}

 public void afterUpdateConnection() {}

 public void beforeUpdateVehicleInfo() {}

 public void tearDown(){

   getLog().infoSimTime(this, "Shutting down road side unit application", new
Object[0]);

 }}
```

## 5.2.2 Castelldefels Vehicle Programming

Vehicles in the Castelldefels simulation are instantiates with two java classes: Slow Down App, and Network Merge Assist.

**Slow Down App:** Defines the slowdown interval of the vehicle. This portion is code is 100% provided by VSimRTI, it is a copy of code used in the Barnim simulation.

**Network Merge Assist:** Network Merge Assist holds code for reading vehicle data for speed, location, route ID, and sensor state. This code equips the vehicle with 802.11p wireless networking and ability to send and receive V2X and DENM message data. Code allowing the vehicle to read environment sensor data and react on sensor readings in a fallback instance are also housed in this java class.

**Vehicle Slow Down Class**

```
package com.telecomsteve.vsimrti.applications;

import ; /*import statements removed for brevity*/
```

```
public class SlowDownApp
 extends AbstractApplication<VehicleOperatingSystem>
 implements VehicleApplication
{
 private boolean hazardousArea = false;
  public SlowDownApp() {}
  public void afterUpdateVehicleInfo() { SensorType[] types =
SensorType.values();

   int strength = 0;
   for (SensorType currentType : types) {
     strength =
((VehicleOperatingSystem)getOperatingSystem()).getStateOfEnvironmentSensor(curren
tType);
     if (strength > 0) {
       break;}}
   if ((strength > 0) && (!hazardousArea)) {

((VehicleOperatingSystem)getOperatingSystem()).changeSpeedWithInterval(6.94444444
4444445D, 5000);
     hazardousArea = true; }
   if ((strength == 0) && (hazardousArea))
   {
     ((VehicleOperatingSystem)getOperatingSystem()).resetSpeed();
     hazardousArea = false; }}
  public void processEvent(Event event)
   throws Exception
 {}
```

```
  public void setUp() {}

  public void tearDown() {}

  public void beforeUpdateConnection() {}

  public void afterUpdateConnection() {}

  public void beforeUpdateVehicleInfo() {}

}
```

**Network Merge Assist Class**

```
package com.telecomsteve.vsimrti.applications;

import ; /* import statements removed for brevity*/


public class NetworkMergeAssist

 extends AbstractApplication<VehicleOperatingSystem>

 implements VehicleApplication, CommunicationApplication

{

 private boolean warned = false;

 private boolean slowed = false;

 private static final int SLOW_DOWN_INTERVAL = 80;

 public NetworkMergeAssist() {}

 public void beforeUpdateConnection() {}

 public void afterUpdateConnection() {}

 public void beforeUpdateVehicleInfo() {}

 public void afterUpdateVehicleInfo() {}

 public void setUp()

 {

   ((VehicleOperatingSystem)getOperatingSystem()).getAdHocModule().enable(new
AdHocModuleConfiguration()

     .addRadio()
```

```
      .channel(AdHocChannel.CCH)

      .power(50)

      .create());

    getLog().infoSimTime(this, "Activated 802.11p AdHoc WiFi Module", new
Object[0]);

    sample();

  }

   private DENM constructDENM()

  {

    GeoCircle geoCircle = new GeoCircle(HAZARD_LOCATION, 3000.0D);

    TopocastDestinationAddress tda =
TopocastDestinationAddress.getBroadcastSingleHop();

    DestinationAddressContainer dac =
DestinationAddressContainer.createTopocastDestinationAddressAdHoc(tda,
AdHocChannel.CCH);

    MessageRouting routing = new MessageRouting(dac,
((VehicleOperatingSystem)getOperatingSystem()).generateSourceAddressContainer());

    InterVehicleMsg message = new InterVehicleMsg(routing,
((VehicleOperatingSystem)getOperatingSystem()).getPosition());

    int strength =
((VehicleOperatingSystem)getOperatingSystem()).getStateOfEnvironmentSensor(Sensor
Type.Speed);


    return new DENM(routing, new
DENMContent(((VehicleOperatingSystem)getOperatingSystem()).getSimulationTime(),
HAZARD_LOCATION, "539695290_211185664_1771967418",

      SensorType.speed, strength, WeatherWarningApp.getNewSpeed(SensorType.speed),
0.0F));

  }

   private void sample() {

    DENM denm = constructDENM();
```

```
((VehicleOperatingSystem)getOperatingSystem()).getAdHocModule().sendV2XMessage(de
nm);
   getLog().infoSimTime(this, "Sent DENM vehicle message", new Object[0]);

((VehicleOperatingSystem)getOperatingSystem()).getEventManager().addEvent(new
Event(((VehicleOperatingSystem)getOperatingSystem()).getSimulationTime() +
2000000000L, this));
 }
  public void tearDown() {
   getLog().infoSimTime(this, "Shutting down Merge Assist application", new
Object[0]);
 }
  public void receiveV2XMessageAcknowledgement(AckV2XMessage ackV2XMessage) {}
 public void beforeGetAndResetUserTaggedValue() {}
 public void afterGetAndResetUserTaggedValue() {}
 public void receiveV2XMessage(ReceivedV2XMessage receivedV2XMessage)
 {
   V2XMessage msg = receivedV2XMessage.getMessage();


   if (!(msg instanceof DENM))
   {return;}


   if
(msg.getRouting().getSourceAddressContainer().getSourceName().equals("rsu_0")) {
     getLog().infoSimTime(this, "Control message received from Roadside Unit",
new Object[0]);
   }
   DENM denm = (DENM)msg;
   getLog().infoSimTime(this, "Processing control message", new Object[0]);
```

```
    getLog().debug("Environment Warning Message Received. Processing...");
  }
   public void processEvent(Event event)
    throws Exception
  {
    if (!isValidStateAndLog()) {
      return; }
    getLog().debugSimTime(this, "Processing event {}", new Object[] { event });
    sample();
    detectSensors();}

  private void detectSensors() {
    SensorType[] types = SensorType.values();
    SensorType type = null;
    int strength = 0;

    for (SensorType currentType : types) {
      strength =
((VehicleOperatingSystem)getOperatingSystem()).getStateOfEnvironmentSensor(curren
tType);

      if (strength > 0) {
        type = currentType;
        break; }}

    if (strength > 0) {
      reactOnEnvironmentData(type, strength);
    } else {
```

```
    getLog().debugSimTime(this, "No Sensor/Event detected", new Object[0]);
  }}

 private void reactOnEnvironmentData(SensorType type, int strength)
 {
   if (((VehicleOperatingSystem)getOperatingSystem()).getVehicleInfo() == null) {
     getLog().infoSimTime(this, "No vehicleInfo given, skipping.", new
Object[0]);
     return; }
   if
(((VehicleOperatingSystem)getOperatingSystem()).getVehicleInfo().getRoadPosition(
) == null) {
     getLog().warnSimTime(this, "No road position given, skip this event", new
Object[0]);
     return;
   }
   float newSpeed = getNewSpeed(type);

   getLog().infoSimTime(this, "Sensored {} event detected, reducing speed to {}
m/s", new Object[] { type, Float.valueOf(newSpeed) });
   getLog().infoSimTime(this, "Position: {}", new Object[] {
((VehicleOperatingSystem)getOperatingSystem()).getPosition() });
   getLog().infoSimTime(this, "SensorType to: {}", new Object[] { type });
   getLog().infoSimTime(this, "CurrVehicle: route: {}", new Object[] {
((VehicleOperatingSystem)getOperatingSystem()).getVehicleInfo().getRouteId() });
   getLog().debugSimTime(this, "Event strength to: {}", new Object[] {
Integer.valueOf(strength) });
   getLog().debugSimTime(this, "RoadId on which the event take place: {}", new
Object[] {
((VehicleOperatingSystem)getOperatingSystem()).getVehicleInfo().getRoadPosition()
.getConnection().getId() });
```

```
   getLog().debugSimTime(this, "CurrVehicle: position: {}", new Object[] {
((VehicleOperatingSystem)getOperatingSystem()).getVehicleInfo().getRoadPosition()
});


   if (!slowed) {

     slowed = true;

     ((VehicleOperatingSystem)getOperatingSystem()).slowDown(newSpeed, 80);

   }}


 public static float getNewSpeed(SensorType type)

 {

   switch (type)

   {

   case Speed:

     return 2.0F;

   case Direction:

     return 8.0F;

   }

   return 10.0F;

 }

  public void beforeSendCAM() {}

  public void afterSendCAM() {}

  public void beforeSendV2XMessage() {}

  public void afterSendV2XMessage() {} }
```

# 6 Results

## 6.1 VSimRTI Code Dependencies

In total twenty-eight import statements were used to pull in various portions of code written by Fraunhofer Institute for VSimRTI. These code blocks represent fundamental function of the simulator, such as message routing and radio medium access control - when using the built-in SNS communications during simulation. A full list of code imports is provided at the end of this section.

During programming, full development of the autonomous merge calculation was prevented by restrictions imposed by the DENM and V2X message formats used to exchange data in VSimRTI. Further research confirms that these controls are required because programming to handle message routing is not limited to vehicles in simulation - but also extends to the synchronization done for the built-in SNS communications, and SUMO routing. Using a java decompiler, I decompiled the .jar file of the Barnim simulation, provided with VSimRTI. This .jar file indeed contains additional class files not provided as part of the sources example code made available to researchers. Within the .jar file, I find code for the InterVehicleMsg class, which defines a string value and is looking explicitly for a sender position value:

```
public final class InterVehicleMsg
 extends V2XMessage{
 private final GeoPoint senderPosition;
 private final EncodedV2XMessage encodedV2XMessage;
 private static final int minLen = 128;
  public InterVehicleMsg(MessageRouting routing, GeoPoint senderPosition)
 {
   super(routing);
   encodedV2XMessage = new EncodedV2XMessage(16, 128);
```

```
    this.senderPosition = senderPosition;
  }
   public GeoPoint getSenderPosition() {
    return senderPosition;
  }
  @NotNull
  public EncodedV2XMessage getEncodedV2XMessage()
  { return encodedV2XMessage;}
   public String toString()
  {
    StringBuffer sb = new StringBuffer("InterVehicleMsg{");
    sb.append("senderPosition=").append(senderPosition);
    sb.append('}');
    return sb.toString();
  }}
```

Modifying the number of arguments InterVehicleMsg accepts causes additional compile errors for code relating to DENM messages, as DENM assumes a format for InterVehicleMsg - which I had since changed. From this moment of dependency error – I complete the Castelldefels simulation code, relying instead on the sensored speed notice for the area surrounding the roundabout.

### Full Dependency List

```
import
com.dcaiti.vsimrti.fed.applicationNT.ambassador.simulationUnit.applicationInterfa
ces.Application;
import
com.dcaiti.vsimrti.fed.applicationNT.ambassador.simulationUnit.applicationInterfa
ces.CommunicationApplication;
```

```
import
com.dcaiti.vsimrti.fed.applicationNT.ambassador.simulationUnit.applicationInterfa
ces.RoadSideUnitApplication;
import
com.dcaiti.vsimrti.fed.applicationNT.ambassador.simulationUnit.applicationInterfa
ces.VehicleApplication;
import
com.dcaiti.vsimrti.fed.applicationNT.ambassador.simulationUnit.applications.Abstr
actApplication;
import
com.dcaiti.vsimrti.fed.applicationNT.ambassador.simulationUnit.communication.AdHo
cModule;
import
com.dcaiti.vsimrti.fed.applicationNT.ambassador.simulationUnit.communication.AdHo
cModuleConfiguration;
import
com.dcaiti.vsimrti.fed.applicationNT.ambassador.simulationUnit.communication.AdHo
cModuleConfiguration.AdHocModuleRadioConfiguration;
import
com.dcaiti.vsimrti.fed.applicationNT.ambassador.simulationUnit.operatingSystem.Ro
adSideUnitOperatingSystem;
import com.dcaiti.vsimrti.fed.applicationNT.ambassador.util.UnitLogger;
import com.dcaiti.vsimrti.rti.enums.SensorType;
import com.dcaiti.vsimrti.rti.eventScheduling.Event;
import com.dcaiti.vsimrti.rti.eventScheduling.EventManager;
import com.dcaiti.vsimrti.rti.geometry.GeoPoint;
import com.dcaiti.vsimrti.rti.interfaces.RandomNumberGenerator;
import com.dcaiti.vsimrti.rti.network.AdHocChannel;
import com.dcaiti.vsimrti.rti.objects.address.DestinationAddressContainer;
import com.dcaiti.vsimrti.rti.objects.address.SourceAddressContainer;
import com.dcaiti.vsimrti.rti.objects.address.TopocastDestinationAddress;
import com.dcaiti.vsimrti.rti.objects.v2x.AckV2XMessage;
```

```
import com.dcaiti.vsimrti.rti.objects.v2x.MessageRouting;

import com.dcaiti.vsimrti.rti.objects.v2x.ReceivedV2XMessage;

import com.dcaiti.vsimrti.rti.objects.v2x.V2XMessage;

import com.dcaiti.vsimrti.rti.objects.v2x.denm.DENM;

import com.dcaiti.vsimrti.rti.objects.v2x.denm.DENMContent;

import java.util.List;

import com.dcaiti.vsimrti.rti.objects.road.IConnection;

import com.dcaiti.vsimrti.rti.objects.road.IRoadPosition;
```

## 6.2 Running the Castelldefels Simulation

After removing incompatible code, I now run the simulation to see logging and message data sent during simulation. Two items are important to note at this step. The first is that, because code relating to messaging of vehicle location, route, and speed are removed; second, the simulation no longer relies on calculation done by the autonomous merge algorithm, rather it uses environment sensors as fallback to slow vehicles entering the target roundabout. Running the Castelldefels simulation is done with the statement below. It includes additional switches at the end to enable visualization using Open Street Maps and assign a brake value of 1 to allow the visualization of traffic in real time, rather than at the higher speed in which the results are calculated.

```
./vsimrti.sh -u [username] -s Castelldefels -v -b 1
```

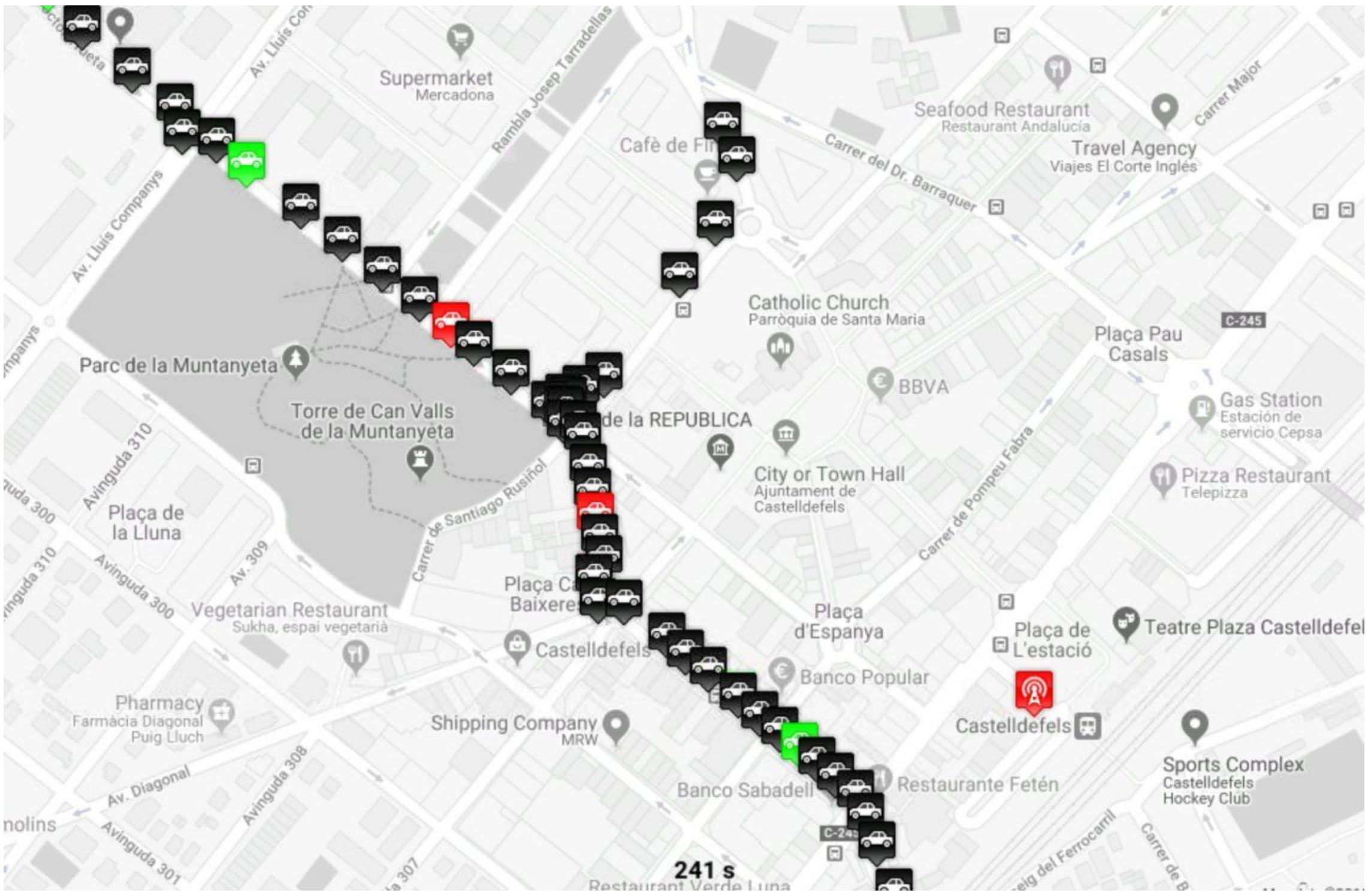

**Figure 11:** Castelldefels vehicle routes, with active 802.11p connectivity

The resulting visualization renders correctly and shows vehicles diverging onto an opposing route, with a final merge happening at the target roundabout. The RSU is now present in the simulation, and red/green coloring signals broadcasts and collisions happening in the ad hoc broadcast range.

During simulation, each entity has a log. All applications mapped to each entity have outputs logged in the "vsimrti/logs" folder. The following section shows output logs from the Castelldefels RSU and Network Merge Assist java classes.

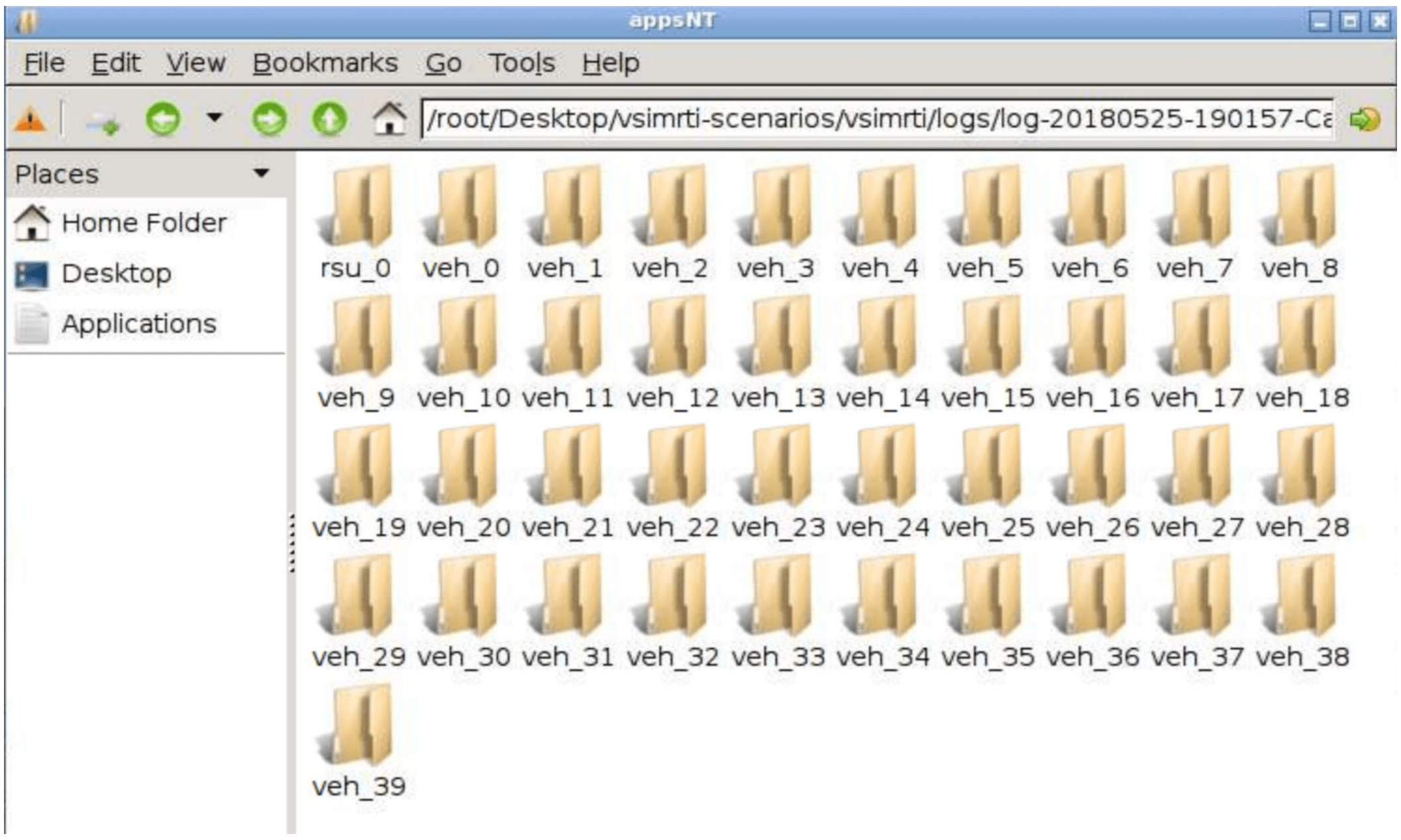

**Figure 12:** VSimRTI Simulation Log Data

### 6.2.1 Roadside Unit Log Messages

The logging output from the Castelldefels RSU java class shows the unit starting 802.11p network connectivity, and it begins receiving broadcast messages from approaching vehicles in range. Important sections of the log are highlighted in blue. Note that as time progresses, additional vehicles arrive in range and the additional broadcasts are interleaved in the RSU logs. The vehicle position data point is explicitly received from each vehicle, as this conforms to the default behavior of the V2X message format used by VSimRTI.

```
2018-07-14 20:21:08,121 INFO  - Starting road side unit application (at
simulation time 0.000,000,000 s)

2018-07-14 20:21:08,130 INFO  - Activating 802.11p AdHoc WiFi Module (at
simulation time 0.000,000,000 s)

2018-07-14 20:21:08,643 INFO  - Received message from veh_0 (at simulation time
91.001,400,000 s)

2018-07-14 20:21:08,643 INFO  - Vehicle position confirmed as:
InterVehicleMsg{senderPosition=GeoPoint{lat|lon=[41.280493,1.974944]}} (at
simulation time 91.001,400,000 s)

2018-07-14 20:21:08,645 INFO  - Sent DENM vehicle message (at simulation time
91.001,400,000 s)
```

2018-07-14 20:21:08,678 INFO  - Received message from veh_0 (at simulation time 96.002,400,000 s)

2018-07-14 20:21:08,678 INFO  - Vehicle position confirmed as: InterVehicleMsg{senderPosition=GeoPoint{lat|lon=[41.280344,1.975151]}} (at simulation time 96.002,400,000 s)

2018-07-14 20:21:08,679 INFO  - Sent DENM vehicle message (at simulation time 96.002,400,000 s)

2018-07-14 20:21:08,693 INFO  - Received message from veh_2 (at simulation time 99.001,400,000 s)

2018-07-14 20:21:08,693 INFO  - Vehicle position confirmed as: InterVehicleMsg{senderPosition=GeoPoint{lat|lon=[41.280524,1.974893]}} (at simulation time 99.001,400,000 s)

2018-07-14 20:21:08,694 INFO  - Sent DENM vehicle message (at simulation time 99.001,400,000 s)

2018-07-14 20:21:08,713 INFO  - Received message from veh_0 (at simulation time 101.000,900,000 s)

2018-07-14 20:21:08,713 INFO  - Vehicle position confirmed as: InterVehicleMsg{senderPosition=GeoPoint{lat|lon=[41.280083,1.975336]}} (at simulation time 101.000,900,000 s)

2018-07-14 20:21:08,714 INFO  - Sent DENM vehicle message (at simulation time 101.000,900,000 s)

2018-07-14 20:21:08,722 INFO  - Received message from veh_1 (at simulation time 102.000,900,000 s)

2018-07-14 20:21:08,722 INFO  - Vehicle position confirmed as: InterVehicleMsg{senderPosition=GeoPoint{lat|lon=[41.281184,1.976052]}} (at simulation time 102.000,900,000 s)

2018-07-14 20:21:08,723 INFO  - Sent DENM vehicle message (at simulation time 102.000,900,000 s)

2018-07-14 20:21:08,732 INFO  - Received message from veh_2 (at simulation time 104.001,400,000 s)

2018-07-14 20:21:08,732 INFO  - Vehicle position confirmed as: InterVehicleMsg{senderPosition=GeoPoint{lat|lon=[41.280397,1.975128]}} (at simulation time 104.001,400,000 s)

2018-07-14 20:21:08,732 INFO  - Sent DENM vehicle message (at simulation time 104.001,400,000 s)

2018-07-14 20:21:08,745 INFO  - Received message from veh_0 (at simulation time 106.000,400,000 s)

2018-07-14 20:21:08,745 INFO  - Vehicle position confirmed as: InterVehicleMsg{senderPosition=GeoPoint{lat|lon=[41.279775,1.975404]}} (at simulation time 106.000,400,000 s)

### 6.2.2 Vehicle Log Messages

Logging data on the vehicles, comes from the Network Merge Assist java class. These log messages show the vehicle also activating 802.11p network connectivity. Details such RoadID and CurrVehicle Position are logged for record, as defined in the logging statements in the application code - but are not forwarded to the RSU. Note that the Position:GeoPoint data *is* sent to the RSU and overlapping timestamps at both the RSU and vehicle confirm the data exchange. Lastly, there is the 'Sensored Speed' event. This confirms that the fallback action is taking place, where the vehicle is notified by road sensors of a speed event, and the vehicle in turn reduces speed to 2 m/s.

```
2018-07-14 20:21:08,190 INFO  - Initialize application (at simulation time
6.000,000,000 s)

2018-07-14 20:21:08,191 INFO  - Activated AdHoc Module (at simulation time
6.000,000,000 s)

2018-07-14 20:21:08,566 INFO  - Sensored Speed event detected, reducing speed to
2.0 m/s (at simulation time 78.000,000,000 s)

2018-07-14 20:21:08,566 INFO  - Position: GeoPoint{lat|lon=[41.281016,1.974083]}
(at simulation time 78.000,000,000 s)

2018-07-14 20:21:08,566 INFO  - Event strength to: 1 (at simulation time
78.000,000,000 s)

2018-07-14 20:21:08,566 INFO  - SensorType to: Speed (at simulation time
78.000,000,000 s)

2018-07-14 20:21:08,566 INFO  - RoadId on which the event take place:
165657372_632874410_632874429 (at simulation time 78.000,000,000 s)

2018-07-14 20:21:08,566 INFO  - CurrVehicle: position:
com.dcaiti.vsimrti.lib.routing.scenariodatabase.road.ScenarioDatabaseRoadPosition
@31180a8a (at simulation time 78.000,000,000 s)

2018-07-14 20:21:08,566 INFO  - CurrVehicle: route: 1 (at simulation time
78.000,000,000 s)

2018-07-14 20:21:08,572 INFO  - Sensored Speed event detected, reducing speed to
2.0 m/s (at simulation time 79.000,000,000 s)

2018-07-14 20:21:08,572 INFO  - Position: GeoPoint{lat|lon=[41.280970,1.974156]}
(at simulation time 79.000,000,000 s)

2018-07-14 20:21:08,573 INFO  - Event strength to: 1 (at simulation time
79.000,000,000 s)
```

```
2018-07-14 20:21:08,573 INFO  - SensorType to: Speed (at simulation time
79.000,000,000 s)

2018-07-14 20:21:08,573 INFO  - RoadId on which the event take place:
165657372_632874410_632874429 (at simulation time 79.000,000,000 s)

2018-07-14 20:21:08,573 INFO  - CurrVehicle: position:
com.dcaiti.vsimrti.lib.routing.scenariodatabase.road.ScenarioDatabaseRoadPosition
@26d66b5b (at simulation time 79.000,000,000 s)

2018-07-14 20:21:08,573 INFO  - CurrVehicle: route: 1 (at simulation time
79.000,000,000 s)

2018-07-14 20:21:08,577 INFO  - Sensored Speed event detected, reducing speed to
2.0 m/s (at simulation time 80.000,000,000 s)

2018-07-14 20:21:08,578 INFO  - Position: GeoPoint{lat|lon=[41.280924,1.974226]}
(at simulation time 80.000,000,000 s)

2018-07-14 20:21:08,578 INFO  - Event strength to: 1 (at simulation time
80.000,000,000 s)

2018-07-14 20:21:08,578 INFO  - SensorType to: Speed (at simulation time
80.000,000,000 s)

2018-07-14 20:21:08,578 INFO  - RoadId on which the event take place:
165657372_632874429_1771967418 (at simulation time 80.000,000,000 s)

2018-07-14 20:21:08,578 INFO  - CurrVehicle: position:
com.dcaiti.vsimrti.lib.routing.scenariodatabase.road.ScenarioDatabaseRoadPosition
@ec1cbc19 (at simulation time 80.000,000,000 s)
```

# 7 Discussion

## 7.1 Conclusion

The aim of this research has been to take a qualitative measure of application layer modeling performance in existing vehicle network simulation platforms. In this paper, I present vehicle network research under Project 5GCar and the Toy Car application under research at Centre Tecnològic de Telecomunicacions de Catalunya (CTTC), before progressing through a history of standards development for vehicle network communications. With this research base, I develop the Castelldefels simulation using a custom autonomous merge algorithm, as a method of benchmarking the capabilities of application layer modeling - in the context of the Cooperative Maneuver use case proposed under Project 5GCar.

Building the Castelldefels simulation required custom programming, and utilized the VSimRTI documented configuration files and API's, while creating a wholly new algorithm that did not directly match an existing pattern defined in IETF, 3GPP, or ITU-R standards. Doing this reveals two important limitation of the VSimRTI platform, being its lengthy closed source code dependencies, as well as a rigid adherence to reference use cases of existing IETF, 3GPP, and ITU-R use cases.

At completion, I successfully modeled vehicle mobility, application simulation, vehicle to vehicle communication and vehicle to infrastructure communication. Mobility controls of the simulation included instantiating varying numbers of vehicles, mapping routes to vehicles, declaring speed, follow distance, braking speed, mapping applications to vehicle, and visualizing the environment using Open Street Maps. Communications within the application layer successfully carried data for environment sensors, and vehicle position. Communications data for vehicle route and speed were successfully logged - but failed to to be integrated and exchanged during simulation, due to code dependency limitations.

# 7.2 Further Research

Although VSimRTI inherits code dependencies, the platform is complete enough that most simulations can be formatted in a way to use only declared VSimRTI java functions. In the case of the autonomous merge algorithm, I propose reworking the algorithm logic, to rely exclusively on reported vehicle position and source container address of V2X messages to deduce the route - instead of attempting to include this data in the V2X message. With the functional application in place, this research can be further extended with accurate modeling of the communications layer using OmNet++ or NS-3, in place of the simpler SNS message exchange functionality built directly into VSimRTI.

## Bibliography

1. Marco Chiani, Andrea Giorgetti, and Enrico Paolini, "Sensor radar for object tracking,"

Proceedings of IEEE, vol. 106, no. 6, pp. 1022-1041, June. 2018

2. Jesus Urena, et al, “Acoustic local positioning with encoded emissions beacons,” Proceedings of IEEE, vol. 106, no. 6, pp. 1042-1062, June. 2018

3. Michael Buehrer, Henk Wymeersch, and Reza Monir Vaghefi, “Collaborative sensor network localization: Algorithms and practical issues,” Proceedings of IEEE, vol. 106, no. 6, pp. 1089-1114, June. 2018

4. Musa Furkan, Ahmet Dundar Sezer, and Sinan Gezici, “Localization via visible light systems,” Proceedings of IEEE, vol. 106, no. 6, pp. 1042-1062, June. 2018.

5. K. Ranta-Aho, "Performance of 3GPP Rel-9 LTE positioning methods," in Proc. 2nd Invitational Workshop Opportunistic RF Localization Next Generation Wireless Devices, Jun. 2010, pp. 1-5.

6. Y. Shang, W. Ruml, Y. Zhang, and M. Fromherz, "Localization from mere connectivity," in Proc. Mobile Ad Hoc Netw. Comput. (MobilHoc), 2003, pp. 201-212.

7. “TheForeignMan”, “DIY Telematics Box”, https://www.instructables.com/id/DIY-Telematics-Box/, 28, May 2018.

8. Image: https://www.plugntrackgps.com/pages/quick-start, Accessed July 2018.

9. Robert Protzmann, Bjorn Schunemann, and Llja Radusch, "The Influences of Communications Models on the Simulated Effectiveness of V2X Applications", IEEE Communications Magazine, pp.149-155, November 2011.

10. Rimon Barr, "JiST - Java in Simulation Time User Guide," http://jist.ece.cornell.edu/docs/040319-jist-user.pdf, pp.1-34, March 19, 2004.

11. Rimon Barr, "SWANS - Scalable Wireless Ad Hoc Network Simulator User Guide," http://jist.ece.cornell.edu/docs/040319-swans-user.pdf, pp.1-15, March 19, 2004.

12. Konstantinos Katsaros et al, "Application of Vehicular Communications for Improving the Efficiency of Traffic in Urban Areas," Special issue on the selected papers of IWCMC 2011, pp.1657-1667, December, 2011.

13. Florian Hausler, Emanuele Crisostomi, Arich Schlote, Llja Radusch, and Robert

Shorten, "Stochastic park-and-charge balancing for fully electric and plug-in hybrid vehicles,". IEEE Transactions on Intelligent Transportation Systems Volume: 15, Issue: 2, April 2014.

14. Charalambos Zinoiou, Konstantinos Katsaros, Ralf Kernchen, Mehrdad Dianati, "Performance Evaluation of an Adaptive Route Change Algorithm Using an Integrated Cooperative ITS Simulation Platform," Conference Proceedings of the Intl Wireless Computing and Mobile Computing Conf (IWCMC). December 2012.

15. Jerome Herri et al, "Modeling and Simulating ITS Applications with iTetris," Proceedings of the 6th ACM workshop on Performance monitoring and measurement of heterogeneous wireless and wired networks. pp. 33-40, 31 October 2011.

16. Antonio Eduardo Fernandez et al, "Deliverable D2.1 5GCAR Scenarios, Use Cases, Requirements and KPIs," https://5gcar.eu/wp-content/uploads/2017/05/5GCAR_D2.1_v1.0.pdf, pp.1-87, 31 August, 2017.

17. 3GPP TR 22.885, "Study on LTE support for Vehicle-to-Everything (V2X) services", 2015.

18. 3GPP TR 22.186, "Service requirements for enhanced V2X scenarios", 2017.

19. ITU-R M.2083-0, "IMT Vision - "Framework and overall objectives of the future development of IMT for 2020 and beyond"", 2015.

20. ITU-R M.1890, "Intelligent Transport Systems - Guidelines and objectives", 2011.

21. Karri Ranta-aho, "Performance of 3GPP Rel-9 LTE positioning methods," Proceedings of the 2nd Opportunistic RF Localization for Next Generation Wireless Devices Conference. 13 June, 2010.

22. ETSI TR 102 638, "Intelligent Transport Systems (ITS); Vehicular Communications; Basic Set of Applications; Definitions", V1.1.1, 2009.

23. ETSI TR 103 298, "Intelligent Transport Systems (ITS); Platooning: pre-standardization study", 2017.

24. ETSI TR 103 299, “Intelligent Transport Systems (ITS); Cooperative Adaptive Cruise Control (C-ACC); Pre-standardization study”, 2017.

25. Bjoern Schuenemann, Llja Radusch, and Kay Massow, “Realistic Simulation of Vehicular Communication and Vehicle-2-X Applications”, Proceedings of the 1st international conference on Simulation tools and technique for communications, networks and systems & workshops, Article No 62, 2008.

26. Cyril Nguyen Van Phu, Nadir Farhi, Habib Haj-Salem, Jean-Patrick Lebacque, “A vehicle-to-infrastructure communication based algorithm for urban traffic control”, Proceedings of the 5th IEEE International Conference on Models and Technologies for Intelligent Transportation Systems (MT-ITS), 2017.

27. Konstantinos Katsaros, Mehrdad Dianati, Karsten Roscher, "A Position-based Routing Module for Simulation of VANETs in NS-3," Proceedings of the 5th international ICST conference on Simulation tools and technique. pp. 343-352, March 2012.

28. Choudhury A., Maszczyk T., Dauwels J., Belagal Math C., Li H., "An Integrated simulation environment for testing V2X protocols and applications," Procedia Computer Science, Vol 80, pp. 2042-2052. 2016.

29. Bjoern Schuenemann, Llja Radusch, "V2X-Based Traffic Congestion Recognition and Avoidance," 10th International Symposium on Pervasive Systems, Algorithms, and Networks, Kaohsiung, pp. 637-641. 2009.

# Appendices

## Appendix 1: Running VSimRTI in a Container

The following section includes further research that was completed as part of experimentation - but were ultimately removed, as they contribute directly to experiment results.

### 1.1 Deploying VSimRTI within Docker

I won't detail the installation of Docker, or management of docker images. It is important to note again that when running VSimRTI for the first time, it is required to have a licensed issued to run simulations. This license is tied to your hardware configuration – fortunately, executing the "firstrun.sh" script of VSimRTI, within Docker; it still sees the native hardware. Simulations can be run within Docker, or directly on the host computer, without a new license.

To pull the docker image:

```
docker pull telecomsteve/vsimrti-web
```

**Running Ubuntu in the Browser**

The first hurtle of making the simulation portable was to get an Ubuntu image running within Docker while having a full desktop UI. This was done using X11, VNC, Apache Server, and LXDE desktop on an Ubuntu 16 LTS core. The origins of this Ubuntu base can be viewed on GitHub. From this core Ubuntu instance additional commands were added to the Docker file to install Java, Git, SUMO, Firefox, and to download, unzip, and place the VSimRTI files on the desktop. When running the Docker image, the included Apache installation and X11 make the Ubuntu LXDE desktop available at port 80 of the Docker machine. This creates a fixed package, that could be run, torn down, and shared between anyone who has simulations to run in VSimRTI.

Once connected to the running Ubuntu desktop, simulation files can be pulled down and run on the Docker machine, using Git. The below screenshot shows the VSimRTI “Tiergarten” example running within Firefox on the Docker Ubuntu image.

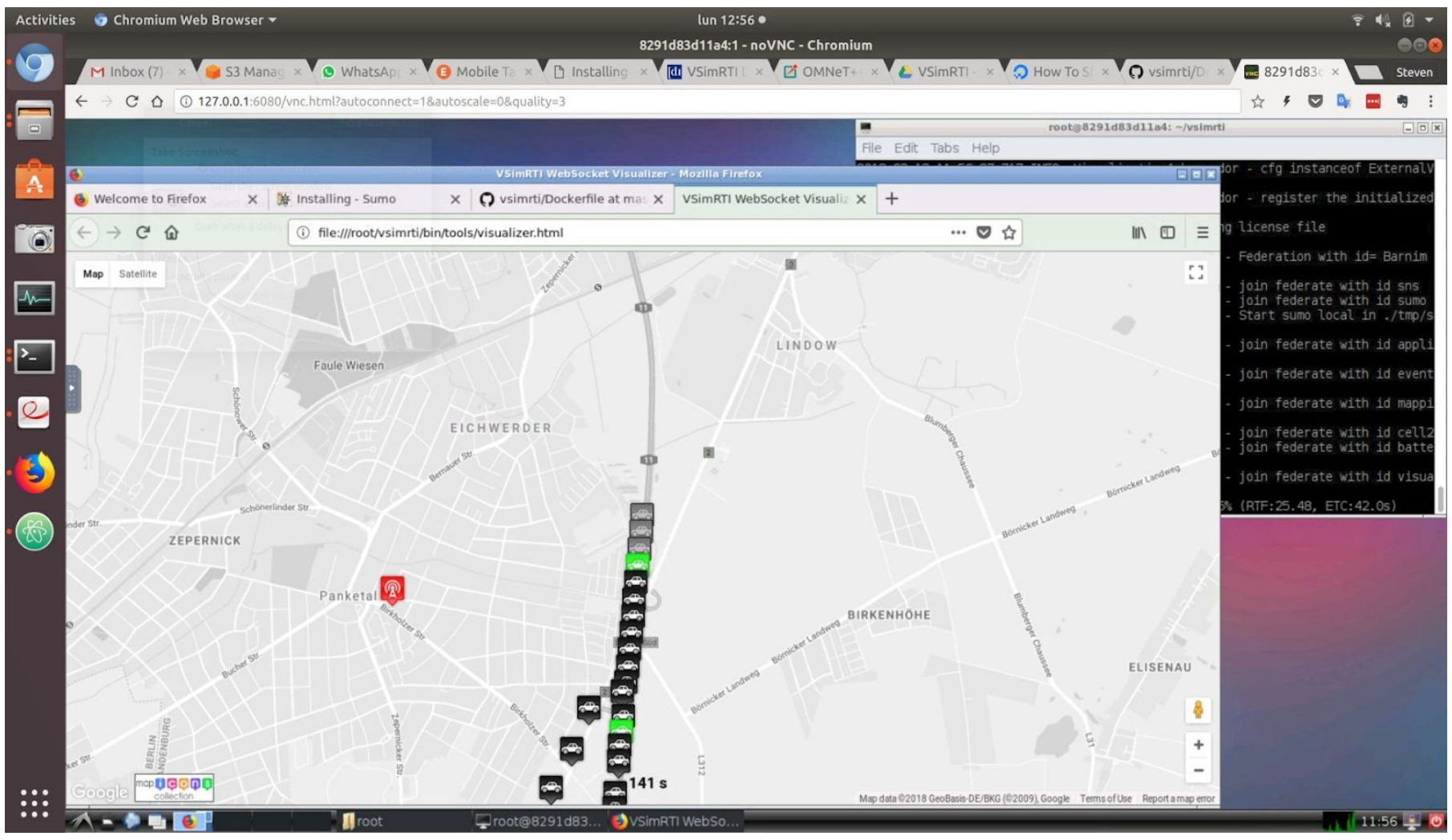

**Figure 13:** VSimRTI running within Docker

To run the image:

```
docker run -it –rm -p 6080:80 telecomsteve/vsimrti-web
```

Access the Ubuntu GUI from the browser:

```
http://127.0.0.1:6080/
```

## Docker Build File

```
FROM ubuntu:16.04
LABEL maintainer="steven@telecomsteve.com"

RUN sed -i 's#http://archive.ubuntu.com/#http://tw.archive.ubuntu.com/#'
/etc/apt/sources.list

# base install packages
RUN apt-get update \
 && apt-get install -y --no-install-recommends software-properties-common \
 && add-apt-repository ppa:fcwu-tw/ppa \
 && add-apt-repository -y ppa:sumo/stable \
```

```
 && apt-add-repository -y ppa:webupd8team/java \
 && apt-get update \
 && apt-get install -y --no-install-recommends --allow-unauthenticated \
 supervisor \
 sudo vim-tiny net-tools lxde x11vnc xvfb python-software-properties
debconf-utils \
 firefox nginx python-pip python-dev build-essential \
 mesa-utils libgl1-mesa-dri dbus-x11 x11-utils \
 dialog wget unzip nano git \
 sumo sumo-tools sumo-doc

# jdk 8 install
RUN echo "oracle-java8-installer shared/accepted-oracle-license-v1-1 select true"
| sudo debconf-set-selections
RUN apt-get install -y oracle-java8-installer

# vsimrti additional packages
RUN wget
https://www.dcaiti.tu-berlin.de/research/simulation/download/get/vsimrti-bin-17.0
.zip
RUN unzip vsimrti-bin-17.0.zip -d /root/Desktop
RUN rm vsimrti-bin-17.0.zip
RUN chmod +x /root/Desktop/vsimrti-allinone/vsimrti/firstStart.sh
RUN bash /root/Desktop/vsimrti-allinone/vsimrti/firstStart.sh

# tini for subreap
ARG TINI_VERSION=v0.9.0
ADD https://github.com/krallin/tini/releases/download/${TINI_VERSION}/tini
/bin/tini
RUN chmod +x /bin/tini

ADD image/usr/lib/web/requirements.txt /tmp/
RUN pip install setuptools wheel && pip install -r /tmp/requirements.txt
ADD image /

# desktop customization
ADD /desktop/panel /tmp/
ADD /desktop/slate.png /tmp/
RUN rm /etc/xdg/lxpanel/default/panels/panel
RUN rm /etc/xdg/lxpanel/LXDE/panels/panel
RUN cp /tmp/panel /etc/xdg/lxpanel/default/panels/panel
RUN cp /tmp/panel /etc/xdg/lxpanel/LXDE/panels/panel

EXPOSE 80
WORKDIR /root
ENV HOME=/home/ubuntu \
 SHELL=/bin/bash
ENTRYPOINT ["/startup.sh"]
```

# Appendix 2: Integrating NS3 and VSimRTI

To get the NS3 build included in the docker container a few additional software is installed, and the NS3 installer shell script is called to completed the NS3 build. These additional Docker build commands are added to the build file reference previously in appendix 1:

```
# base install packages
RUN apt-get update \
 && apt-get install -y --no-install-recommends software-properties-common \
 && add-apt-repository ppa:fcwu-tw/ppa \
 && add-apt-repository -y ppa:sumo/stable \
 && apt-add-repository -y ppa:webupd8team/java \
 && apt-get update \
 && apt-get install -y --no-install-recommends --allow-unauthenticated \
 supervisor \
 sudo vim-tiny net-tools lxde x11vnc xvfb python-software-properties
debconf-utils \
 firefox nginx python-pip python-dev build-essential \
 mesa-utils libgl1-mesa-dri dbus-x11 x11-utils \
 dialog wget unzip nano git libprotobuf-dev gcc g++ bison flex lbzip2 libxml2-dev
rsync libsqlite3-dev patch


# vsimrti additional packages
RUN git clone https://github.com/stevenplatt/vsimrti-scenarios.git
/root/Desktop/upf/

# NS3 install
RUN yes "y" | bash /root/Desktop/upf/vsimrti/bin/fed/ns3/ns3_installer.sh
```

## 2.2 Detailed Logging with NS3

After bundling NS3 into the Docker build, detailed communications layer logging was enabled by modifying the *vsimrti/etc/logback.xml* file, to set NS3 logging to be enabled, with log level of "DEBUG". Finally, the *vsimrti_config.xml* file within each scenario folder was updated to enable use of NS3 for communications in the simulation, and to disable the default enabled 'SNS' communications simulator. Doing this allowed using propagation models from NS3 and get full communication logging as shown below.

**Logging Before:**

```
2018-04-12 16:29:13,270 INFO  SnsAmbassador - Send v2xMessage.id=19 from
node=rsu_0 as Topocast (singlehop) @time=18.000,000,000 s

2018-04-12 16:29:13,271 INFO  SnsAmbassador - Receive v2xMessage.id=19 on
node=veh_0 @time=18.000,900,000 s

2018-04-12 16:29:13,271 INFO  SnsAmbassador - Receive v2xMessage.id=19 on
node=veh_1 @time=18.000,900,000 s

2018-04-12 16:29:13,271 INFO  SnsAmbassador - Receive v2xMessage.id=19 on
node=tl_0 @time=18.000,900,000 s

2018-04-12 16:29:13,274 INFO  SnsAmbassador - Send v2xMessage.id=20 from=veh_2 as
Geocast (geo routing) @time=19.000,000,000 s
```

**Logging After:**

```
2018-04-12 16:35:15,786 DEBUG Ns3Ambassador - ProcessMessage VehicleMovements at
t=75000000000

2018-04-12 16:35:15,786 DEBUG Ns3Ambassador - Update Vehicle Positions

2018-04-12 16:35:15,786 DEBUG Ns3Ambassador - UpdateNode : ID[int=veh_2, ext=3]

2018-04-12 16:35:15,786 DEBUG Ns3Ambassador - Pos: x(918.2701732605929)
y(173.16697777435184) Point2D.Double: GeoPoint{lat|lon=[52.513036,13.330212]}

2018-04-12 16:35:15,786 DEBUG Ns3Ambassador - UpdateNode : ID[int=veh_3, ext=4]
```

```
2018-04-12 16:35:15,786 DEBUG Ns3Ambassador - Pos: x(816.3387485357234)
y(162.39611352980137) Point2D.Double: GeoPoint{lat|lon=[52.512918,13.328714]}

2018-04-12 16:35:15,786 DEBUG Ns3Ambassador - UpdateNode : ID[int=veh_0, ext=1]

2018-04-12 16:35:15,786 DEBUG Ns3Ambassador - Pos: x(903.0686549004749)
y(174.82249094638973) Point2D.Double: GeoPoint{lat|lon=[52.513047,13.329987]}

2018-04-12 16:35:15,786 DEBUG Ns3Ambassador - UpdateNode : ID[int=veh_1, ext=2]

2018-04-12 16:35:15,786 DEBUG Ns3Ambassador - Pos: x(865.8905054948991)
y(167.60215506237) Point2D.Double: GeoPoint{lat|lon=[52.512975,13.329442]}

2018-04-12 16:35:15,787 DEBUG Ns3Ambassador - ProcessTimeAdvanceGrant at
t=75000000000

2018-04-12 16:35:15,787 DEBUG Ns3Ambassador - Requested next_event at 75000000000
```